\documentclass[aps,preprint,superscriptaddress,nofootinbib]{revtex4-1}

\usepackage{amsmath,amssymb,amsfonts,bm, slashed,graphicx,longtable}
\usepackage{array,multirow}
\usepackage{hyperref}

\newcommand{\amH}[3]{\big\langle #1 \big| #2 \big| #3 \big\rangle}

\newcommand{\Acp}[2]{A_{\rm CP}[#1 \to #2]}
\newcommand{\GAcp}[2]{\delta_{\rm CP}[#1 \to #2]}
\newcommand{\gAcp}{\delta_{\rm CP}}
\newcommand{\DAcp}[2]{\Delta_{\rm CP}[#1 \to #2]}
\newcommand{\dAcp}{\Delta_{\rm CP}}

\newcommand{\dr}[2]{\bar{\Gamma}[#1 \to #2]}
\newcommand{\ps}[2]{\mathcal{P}_{(#1; #2)}}

\def\lqcd{\Lambda_{\rm QCD}}
\def\ov{\overline}
\newcommand{\nn}{\nonumber}
\newcommand{\wt}{\widetilde}

\newcounter{incre}
\newcommand{\benum}{\begin{enumerate}\setcounter{enumi}{\value{incre}}}
\newcommand{\eenum}{\setcounter{incre}{\value{enumi}}\end{enumerate}}

\usepackage{xcolor}
\definecolor{red}{rgb}{1.0, 0, 0}
\definecolor{blue}{rgb}{0, 0, 1.0}

\allowdisplaybreaks[1]

\makeatletter
\newcommand{\oset}[2]{%
  {\mathop{#2}\limits^{\vbox to -.5\ex@{\kern-\tw@\ex@
   \hbox{\scriptsize #1}\vss}}}}
\makeatother

\begin{document}

\title{\boldmath
More Flavor SU(3) Tests for New Physics in CP Violating \texorpdfstring{$B$}{B} Decays}

\author{Yuval Grossman}
\email{yg73@cornell.edu}
\affiliation{Laboratory for Elementary-Particle Physics, Cornell University, Ithaca, N.Y.}

\author{Zoltan Ligeti}
\email{ligeti@lbl.gov}
\affiliation{Ernest Orlando Lawrence Berkeley National Laboratory,
University of California, Berkeley, CA 94720}

\author{Dean J.\ Robinson}
\email{djrobinson@berkeley.edu}
\affiliation{Laboratory for Elementary-Particle Physics, Cornell University, Ithaca, N.Y.}
\affiliation{Ernest Orlando Lawrence Berkeley National Laboratory,
University of California, Berkeley, CA 94720}
\affiliation{Department of Physics, University of California, Berkeley, CA 94720, USA}

\begin{abstract}

The recent LHCb measurements of the $B_s \to K^-\pi^+$ and $B_s \to K^+K^-$
rates and CP asymmetries are in agreement with U-spin expectations from $B_d \to
K^+\pi^-$ and $B_d \to \pi^+\pi^-$ results. We derive the complete set of
isospin, U-spin, and SU(3) relations among the CP asymmetries in two-body
charmless $B \to PP$ and $B \to PV$ decays, some of which are novel. To go
beyond the unbroken SU(3) limit, we present relations which are properly defined
and normalized to allow incorporation of SU(3) breaking in the simplest manner.
We show that there are no CP relations beyond first order in SU(3) and isospin
breaking. We also consider the corresponding relations for charm decays.
Comparing parametrizations of the leading order sum rules with data can shed
light on the applicability and limitations of both the flavor
symmetry and factorization-based descriptions of SU(3) breaking. Two
factorization relations can already be tested, and we show they agree with
current data.

\end{abstract}

\maketitle

\section{Introduction}

One of the interesting open questions related to the Belle~\cite{Lin:2008zzaa}
and BaBar~\cite{Aubert:2007mj} data is the interpretation of the difference of
CP asymmetries, $\Acp{B^+}{K^+\pi^0} - \Acp{B^0}{K^+\pi^-} = 0.126 \pm
0.022$~\cite{*[] [{, and updates at
\url{http://www.slac.stanford.edu/xorg/hfag/}.}] Amhis:2012bh}.  This
measurement is in tension with the results of calculations using an expansion
about the heavy quark limit $m_b \gg \lqcd$~\cite{Beneke:2001ev, Bauer:2004tj,
Keum:2000wi} (see, however,~\cite{Li:2005kt, Bai:2013tsa}). In contrast, the
data do not present difficulties if flavor SU(3) symmetry is the only
theoretical input used to relate the direct CP asymmetries. 

We can hope to achieve a better understanding of the applicability and
limitations of these approaches by exploring other relations for CP violation
among charmless, two-body $B$ decays. Such an understanding would not only
enhance the ability of future $B$ decay measurements to probe for new physics
(NP) signals, but also improve our understanding of QCD.  For example, the
failure of an SU(3) relation at a larger than expected level may be due to a NP
signal, and could tell us about the flavor structure of NP.  Alternatively, if
predictions of factorization fail, then undertanding as well as possible under
what circumstances that occurs may in turn improve our understanding of the QCD
dynamics. For example, one might learn that the relative strong phase of the
so-called tree and color-suppressed tree amplitudes in the diagrammatic picture
is large in some cases, despite being power suppressed in the heavy quark limit.

With these motivations in mind, the LHCb Collaboration has recently reported the first evidence of CP violation
(CPV) in $B_s \to K^- \pi^+$ decay~\cite{Aaij:2013iua}. This observation has been
combined with existing data for $B_d \to K^+\pi^-$~\cite{pdg:2012} to probe the
SM through the parameter $\Delta$~\cite{He:1998rq, Gronau:2000zy,
Lipkin:2005pb}, for which the result is quoted as~\cite{Aaij:2013iua}
\begin{equation}
	\label{eqn:DEV}
	\Delta \equiv \frac{\Acp{B_d }{K^+\pi^-}}{\Acp{B_s }{ K^-\pi^+}}
	+ \frac{\dr{B_s }{ K^-\pi^+}}{\dr{B_d }{ K^+\pi^-}}
	= -0.02 \pm 0.05 \pm 0.04~.
\end{equation}
Here the experimentally measured direct CP asymmetries, $A_{\rm CP}$,
are defined to be
\begin{equation}
\Acp{i}{f} \equiv \frac{\DAcp{i}{f}}{2\,\dr{i}{f}}\,, 
\end{equation}
where the initial state, $i$, is conventionally~\cite{pdg:2012, Amhis:2012bh} a
$B$ meson containing a $\bar{b}$ quark, and
\begin{equation}\label{Gamdef}
\DAcp{i}{f} \equiv \Gamma\big(\bar i\to\bar f\big) - \Gamma\big(i\to f\big)~, \qquad
\dr{i}{f} \equiv \frac12\, \Big[ \Gamma\big(i\to f\big)
  + \Gamma\big(\bar i\to\bar f\big) \Big]~.
\end{equation}
In the SU(3) limit $\Delta=0$, and thus a measurement of $\Delta$ that deviates
significantly from zero may indicate the presence of new physics. For example,
such a deviation may arise from  enhanced contributions to electroweak
penguins. 

While the present experimental result is consistent with zero, one should also
expect deviations from $\Delta = 0$ due to SU(3) breaking effects. The typical
expected size of SU(3) breaking at the amplitude level, parametrized by  $\varepsilon$, is of order
$(m_s - m_d)/\Lambda_{\rm QCD}$ or $f_K/f_\pi -1$, both of which are
$\mathcal{O}(20\%)$. However, for relations between squared amplitudes, such as $\Delta$, the typical SU(3) breaking should be $2\varepsilon$, the factor of two arising from the Taylor expansion in $\varepsilon$. Taking into account an additional suppression factor of about 4, one expects $\Delta \sim 10\%$, in good agreement with the data.  This
suppression factor arises from a ratio of decay rates prefactor, which is a consequence of the definition of $\Delta$.

In order to examine SU(3) breaking effects, parameters such as $\Delta$ are
poorly defined, as they not only carry an arbitrary normalization, but
also unnecessarily introduce SU(3) breaking from phase space. A
more suitable parameter for the study of SU(3) breaking is the properly
normalized and defined combination of these rates and asymmetries,
\begin{equation} 
	\label{eqn:DTD}
	\wt\Delta \equiv \frac{\GAcp{B_d}{K^+\pi^-} +
	\GAcp{B_s}{K^-\pi^+}}{\GAcp{B_d}{K^+\pi^-} - \GAcp{B_s}{K^-\pi^+}} 
	= 0.026 \pm 0.106\,,
\end{equation}
where we defined
\begin{equation}
	\label{eqn:GDR}
	\GAcp{i}{f}= 8\pi\; \ps{i}{f}\, \DAcp{i}{f}\,, \qquad
	\ps{i}{f} \equiv \frac{m_i^2}{|\vec p_{i\to f}|} \,,
\end{equation}
and the data we used is collected in Table~\ref{values} in the main text.
Here $\gAcp$ is the asymmetry of the squared amplitudes for the CP conjugate
decays, that is obtained by removing the $i \to f$ phase space factor,
$1/\ps{i}{f}$, from the rates. The $\vec p_{i \to f}$ is the center of mass
three-momentum of the final state particles; and $m_i$ denotes the initial $B$
meson mass. It is advantageous to use $\wt\Delta$-like parameters instead of
$\Delta$ to parametrize SU(3) breaking, and we consequently express all SU(3)
relations in terms of $\gAcp$ rather than $\dAcp$. The numerical result in
Eq.~(\ref{eqn:DTD}), obtained using the most recent experimental
data~\cite{Amhis:2012bh, Aaij:2013iua}, is in agreement with the na\"\i vely
expected size of SU(3) breaking,~$\varepsilon$.

In Refs.~\cite{He:1998rq, Gronau:2000zy} five other U-spin relations were
presented, which lead to testable parameters similar to $\Delta$ or $\wt
\Delta$. Recently, Ref.~\cite{He:2013vta} presented some  SU(3) CP relations for
$B \to PV$ decays ($P$ denotes a pseudoscalar and $V$ a vector meson). In the
present paper, we extend the results of Refs.~\cite{He:1998rq, Gronau:2000zy, Grossman:2003qp, Nagashima:2007qn,
He:2013vta}, by presenting the full set of SU(3) relations in terms of $\gAcp$
for mesons in both the mass and flavor basis, that is, with and without
octet-singlet and neutral $K$ meson mixing. We consider both $B \to PP$ and $B
\to PV$ decays. 

We further look for relations that hold to second order in SU(3) and isospin
breaking by the quark mass spurion. We show that in the flavor basis, apart from isospin relations, there
exists one CP relation for $B \to PP$ that holds beyond first order SU(3)
breaking, and that this relation also holds beyond first order in isospin
breaking. Once octet-singlet mixing is included, we find there exist no CP
relations beyond first order in SU(3) breaking for either $B \to PP$ or $B \to
PV$, with the exception of isospin relations.  In our analyses, we only consider
effects that are first order in the weak interaction (e.g. we neglect
electroweak penguins), since SU(3) breaking arising from the quark mass spurion is expected to be much larger
than higher-order weak interaction effects.

In parallel to this analysis, we apply QCD and SCET factorization to
study SU(3) breaking effects. In this approach, we may derive relations between
different parameters that vanish in the SU(3) limit. At present there are two such relations that can be tested, and we show that they are in
agreement with the currently available data.

This paper is structured as follows. We first recapitulate the U-spin analysis,
using the more compact Wigner-Eckart picture, and proceed to consider the
effects of U-spin breaking by the strange quark mass. We then present
$\Delta$-type parameters for the charged mesons, and introduce natural,
well-defined parameters for the characterization of U-spin breaking in CP
relations. We derive factorization-based relations between some of these
parameters and compare with current data where possible. Finally, we present the full set of SU(3) relations for both $B \to
PP$ and $B \to PV$ decays. Similar relations that hold for the $D$ meson decays
are presented in an Appendix.

\section{Group Theoretic Analysis}
\label{sec:GTA}

\subsection{CP Sum Rules}
\label{sec:CPSR}

Let us first derive the Wigner-Eckart decomposition for direct CP asymmetries in
the general group theoretic case. This decomposition is well-suited for
expansions in symmetry breaking parameters. 

We are interested in matrix elements of the form
\begin{equation}
\mathcal{A}_{\mu \to \alpha\beta} \equiv \amH{P_\alpha P_\beta}{H}{B_\mu},
\qquad
\ov{\mathcal A}_{\mu \to \alpha\beta} \equiv 
  \amH{\overline P_\alpha \overline P_\beta}{H}{\overline B_\mu},
\end{equation}
where $P_{\alpha}$ denotes the final state mesons, $B_\mu$ is the initial state
and $H$ is the effective Hamiltonian.  The Wigner-Eckart theorem ensures that we
can write these amplitudes in terms of reduced matrix elements, 
\begin{equation}
	\label{eqn:WEDA}
	\mathcal{A}_{\mu \to \alpha\beta} =  \sum_w X_w\, 
	\partial_{P_\alpha P_\beta B_\mu}\, I_w~,
\end{equation}
where $X_w$ are reduced matrix elements, and $I_w$ are group theoretic
invariants, formed from the effective Hamiltonian, initial and final state
tensors. In general, $I_w$ contain both strong and weak phases arising from the
effective Hamiltonian, so it is convenient to write, without loss of
generality, 
\begin{equation}
	\partial_{P_\alpha P_\beta B_\mu}\, I_w \equiv 
	\sum_q \chi^q_{w,\alpha\beta\mu}\, \exp\{i \sigma^q_w\}~, 
\end{equation}	
where $\chi^q_{w,\alpha\beta\mu}$ contain weak phases and group theoretic
coefficients that depend on the particular initial and final states and the
effective Hamiltonian, while $\sigma^q_w$ are strong phases from the effective
Hamiltonian alone. 

The corresponding decay rate for each process is
\begin{equation}
	\label{eqn:DRD}
	\Gamma[B_{\mu} \to P_\alpha P_\beta] = \frac1{8\pi}\,
	\frac1{\ps{\mu}{\alpha\beta}}\, 
	\big|\mathcal{A}_{\mu \to \alpha\beta}\big|^2\times
	\Bigg\{ \begin{matrix}
	1\,, & ~~P_\alpha \neq P_\beta\,, \\ 1/2\,, & ~~P_\alpha = P_\beta\,,
	\end{matrix}
\end{equation}
where $m_{B_\mu}$ is the mass of the initial $B$ meson.  The symmetry factor
$1/2$ arises when the two final state particles are identical, which will be
relevant in Sec.~\ref{sec:SU3R}.  We are interested in relations involving the
difference of CP conjugate square amplitudes, $\gAcp$, which are pure group
theoretic objects: they do not involve phase space factors. That is, we seek sum
rules among
\begin{equation}
	\label{eqn:GA}
	\GAcp{B_\mu}{P_\alpha P_\beta} = \Big( 
	\big|\overline{\mathcal{A}}_{\mu \to \alpha\beta}\big|^2
	- \big|\mathcal{A}_{\mu \to \alpha\beta}\big|^2 \Big) \times
	\Bigg\{ \begin{matrix}
	1\,, & ~~P_\alpha \neq P_\beta\,, \\ 1/2\,, & ~~P_\alpha = P_\beta\,.
	\end{matrix}
\end{equation}
Dropping explicit inclusion of the symmetry factor of $1/2$, we then have
\begin{align}
	\GAcp{B_\mu}{P_\alpha P_\beta}  
	& = \sum_{w,v} \Big( X_w\, \partial_{P_\alpha P_\beta B_\mu}\,
	  \overline{I}_w\, X_v^*\, \partial_{P_\alpha P_\beta B_\mu}\,
	  \overline{I}_v^* - X_w\,\partial_{P_\alpha P_\beta B_\mu}\, I_w\, 
	  X_v^*\, \partial_{P_\alpha P_\beta B_\mu} I_v^*\Big)\notag\\
	& = \sum_{w,v; q,r} \!\!X_wX_v^* \exp\{i (\sigma^q_w- \sigma^r_v)\}\Big[ \chi^{q*}_{w,\alpha\beta\mu}~\chi^{r}_{v,\alpha\beta\mu} - \chi^q_{w,\alpha\beta\mu}~\chi^{r*}_{v,\alpha\beta\mu} \Big]\notag\\
	& = 4 \sum_{w,v; ~q\le r} 2^{-\delta_{qr}}\mbox{Im}\Big[X^*_wX_v\exp\{i (\sigma^r_v- \sigma^q_w ) \}\Big] \mbox{Im}\Big[\chi^{q*}_{w,\alpha\beta\mu}~\chi^{r}_{v,\alpha\beta\mu} \Big]~. \label{eqn:CPGR}
\end{align}

Now, a CP sum rule is a symbol -- i.e. an array of numerical coefficients -- $\mathcal{S}$ such that
\begin{equation}
	\mathcal{S}^{\alpha\beta\mu}\, \GAcp{B_\mu}{P_\alpha P_\beta} = 0~.
\end{equation}
It follows from Eq.~\eqref{eqn:CPGR} that a sufficient condition for sum rules is
\begin{equation}
	\label{eqn:SRGR}
	\mathcal{S}^{\alpha\beta\mu}\, \mbox{Im}\Big[\chi^{q*}_{w,\alpha\beta\mu}~\chi^{r}_{v,\alpha\beta\mu} \Big] = 0~.
\end{equation}
That is, one needs only compute the kernel of
$\chi^{q*}_{w,\alpha\beta\mu}~\chi^{r}_{v,\alpha\beta\mu} $ with respect to the
basis of modes, indexed by $\alpha\beta\mu$. The structure of
$\chi^{q*}_{w,\alpha\beta\mu}~\chi^{r}_{v,\alpha\beta\mu} $ is determined in
part by the group theoretic indices $w,v$, which encode the group theoretic
structure of the initial, final states and effective Hamiltonian. It is further
determined by the strong phase indices $q,r$, which encode the strong phase
structure of the effective Hamiltonian. However, the sum rules are independent
from particular values of these strong phases. Moreover, the sum rules are
independent from the reduced matrix elements $X_w$, and consequently any strong
phase structure carried by these.

In Eq.~\eqref{eqn:CPGR}, we also see that if the amplitude \eqref{eqn:WEDA} for
a CP violating mode involves $n$ invariants, then the corresponding $\gAcp$
involves $n^2$. It is therefore reasonable to expect that it is more difficult
to obtain CP relations, compared with amplitude relations.

\subsection{U-spin Analysis}
\label{sec:UA}

We may now present a compact recapitulation of the derivation of $\Delta$ and
other similar relations in the U-spin limit. Our results agree with those of
\cite{He:1998rq,Gronau:2000zy}, but are presented in a different way.  In the
remainder of this section we only consider decays into a pair of charged
pseudoscalars, so that $P_{\alpha} = (K,\pi)$ are the charged kaon or pion
final states and $B_\mu = (B_d,B_s)$ is the initial state.

First, the neutral $B$ mesons furnish a U-spin anti-doublet
\begin{equation}
[B]_i = \begin{pmatrix} B_d & B_s \end{pmatrix}\;,
\end{equation}
while the two-particle final states furnish singlet and triplet U-spin
representations,
\begin{equation}\label{KpiUspin}
	[M_0] =  \dfrac{\pi^+\pi^- + K^+K^-}{2}~, \qquad
	[M_1]^i_j = \!\begin{pmatrix} \dfrac{\pi^+\pi^- - K^+K^-}{2} & \pi^-K^+ \\ 
	  \pi^+ K^- & \dfrac{K^+K^- - \pi^+\pi^- }{2}\end{pmatrix}.
\end{equation}
Next, the Hamiltonian is a $\Delta U=1/2$ operator. Using CKM unitarity it can be
written in its most general form as
\begin{equation}
\label{eqn:HUR}
H = H^{\rm t} + H^{\rm p}\,,\qquad
[H^{\rm t}]^j = \mathcal{T} \begin{pmatrix}
	V^{\phantom{*}}_{ ud} V^*_{u b}  \\
	V^{\phantom{*}}_{us} V^*_{u b} \end{pmatrix} \equiv \mathcal{T}\lambda^j_u~, \qquad
[H^{\rm p}]^j = \mathcal{P} \begin{pmatrix}
	V^{\phantom{*}}_{cd} V^*_{c b} \\
	V^{\phantom{*}}_{ cs} V^*_{c b} \end{pmatrix}  \equiv \mathcal{P}\lambda^j_c\,. 
\end{equation}
Here $\lambda_q^j \equiv V_{q j} V^*_{q b}$ carry weak phases, and $\mathcal{T}$
and $\mathcal{P}$ are complex numbers containing strong phases.  While the
notation $\mathcal{T}$ and $\mathcal{P}$ is suggestive of `tree' and `penguin',
we emphasize that certain penguins with the same weak phase and flavor
transformation properties as the trees have been absorbed into $H^{\rm t}$.

The Wigner-Eckart theorem ensures that we can write the amplitudes in terms of
two reduced matrix elements in the U-spin limit (cf., Eq.~\eqref{eqn:WEDA}),
\begin{equation}
	\label{eqn:WEA}
	\mathcal{A}_{\mu \to \alpha\beta} \equiv \amH{P_\alpha
	P_\beta}{H}{B_\mu} = \frac{\partial^3}{\partial P_\alpha \partial
	P_\beta \partial B_\mu}\, \Big\{X_1 [M_1]^i_j [B]_i H^j + X_0 [M_0] [B]_iH^i\Big\}\,,
\end{equation}
where the summations over tensor indicies $i$ and $j$ are implicit. In the
present U-spin case, since the Hamiltonian has only $\Delta U = 1/2$ terms,
we can further partition the amplitudes into the form
\begin{equation}
	\label{eqn:AFP}
	\mathcal{A}_{\mu \to \alpha\beta} = \sum_w X_w [C_{w,j}]_{\alpha\beta\mu} H^j~,
\end{equation}
where we have defined the following 
\begin{equation}
	\label{eqn:LOI}
	[C_{1,j}]_{\alpha\beta\mu} \equiv \partial_{P_\alpha P_\beta B_\mu}
	\Big\{ [M_1]^k_j [B]_k\Big\} \,,\qquad 
	[C_{0,j}]_{\alpha\beta\mu} \equiv \partial_{P_\alpha P_\beta B_\mu} 
	\Big\{ [M_0] [B]_j\Big\} \,,
\end{equation}
which we hereafter refer to as `partial invariants', since they form part of the group theoretic invariants.
Note for all charged meson final state and $B$ initial state combinations, the
partial invariants happen to have the property that
\begin{equation}
	\label{eqn:CCR}
	[C_{w,i}][C_{v,j}] = 0~, \quad i \not= j~.
\end{equation}

We now have from Eqs.~\eqref{eqn:HUR} and \eqref{eqn:AFP},
\begin{equation}
	\partial_{\alpha\beta\mu}I_w =  [C_{w,j}]_{\alpha\beta\mu} \big[\mathcal{T}\lambda_{\rm u}^j + \mathcal{P}\lambda_{\rm c}^j\big]~,
\end{equation}
with $w = 0,1$ in the U-spin limit. Applying Eq.~\eqref{eqn:CPGR}, we have
$\chi^{1,2}_w = |\mathcal{T}|\lambda_{\rm u}^jC_{w,j}$\,,
$|\mathcal{P}|\lambda^j_{\rm c}C_{w,j}$\,, and $\sigma^{1,2}_w =
\mbox{arg}[\mathcal{T}]$, $\mbox{arg}[\mathcal{P}]$ so it immediately follows
that
\begin{equation}
	\label{eqn:WEUD}
	\GAcp{B_\mu}{P_\alpha P_\beta}  = 4\, |\mathcal{P\, T}|\,
	\sum_{w,v } \mbox{Im}\Big[X^*_wX_ve^{i \delta}\Big][\mathcal{M}_{\rm CP}]_{w,v; \alpha\beta\mu}~,
\end{equation}
where $\delta \equiv \mbox{arg}[\mathcal{P\, T}^*]$, and we defined
\begin{equation}
	[\mathcal{M}_{\rm CP}]_{w,v; \alpha\beta\mu} \equiv  [C_{w,j}]_{\alpha\beta\mu}  [C_{v,k}]_{\alpha\beta\mu}  \mbox{Im}\big[\lambda_{\rm u}^{j*}\lambda_{\rm c}^k\big]~.
\end{equation}
Contributions to $\gAcp$ are generated by interference between terms
which carry a relative weak and strong phase. Here, the only such interference
terms are cross terms between $\lambda^j_{u}$ and $\lambda^k_{c}$, which is
precisely the term that appears in Eq.~\eqref{eqn:WEUD}. Contributions such as
$\mbox{Im} [\lambda_{\rm u}^{j*}\lambda_{\rm u}^k]$ or $\mbox{Im} [\lambda_{\rm
c}^{j*}\lambda_{\rm c}^k]$ do not occur because Eq.~\eqref{eqn:CCR} enforces $j
=k$, so these imaginary parts are zero.

Explicitly, the operator that generates CP violation
\begin{equation}
	\label{eqn:OCP}
	\big[\mathcal{O}_{\rm CP}\big]^{ij}  \equiv \mbox{Im}\big\{\lambda_{\rm u}^{*i} \lambda_{\rm c}^{j}\big\} = 
	\begin{pmatrix} \mbox{Im}[V^{\phantom{*}}_{cd} V^*_{cb} V^*_{ud} V^{\phantom{*}}_{ ub}] 
	& \quad \mbox{Im}[V^{\phantom{*}}_{cd} V^*_{cb} V^*_{us} V^{\phantom{*}}_{ub}] \\ 
	\mbox{Im}[V^{\phantom{*}}_{cs} V^*_{cb} V^*_{ud} V^{\phantom{*}}_{ub}] 
	& \quad \mbox{Im}[V^{\phantom{*}}_{cs} V^*_{cb} V^*_{us} V^{\phantom{*}}_{ub}] \end{pmatrix}  \equiv \begin{pmatrix} J & \ldots \\ \ldots & -J \end{pmatrix} ~,
\end{equation}
where $J$ is the Jarlskog invariant. In the notation of Eq.~\eqref{eqn:WEUD},
the CP asymmetry for each mode has now been partitioned into reduced matrix
elements, $X_w$, partial invariants $[C_w]_{\alpha \beta\mu}$ that depend on the
group theoretic structure of the initial and final states and the Hamiltonian,
and a CP operator $\mathcal{O}_{\rm CP}$, that arises from the CKM structure of
the Hamiltonian alone. We emphasize that the subscripts $i$ and $j$ are
implicitly summed tensor indices, the indices $w,v$ label the different possible
partial invariants, while $\alpha,\beta,\mu$ label the initial and final
states.  For the U-spin representations under consideration, in the U-spin limit
the partial invariants are specified in Eq.~\eqref{eqn:LOI}. 

Note that the off-diagonal terms of $\mathcal{O}_{\rm CP}$ are basis dependent,
while the diagonal, Jarlskog, terms are independent of the choice of up-type
quark basis for the Hamiltonian. That is, they are independent of which term is
chosen to be eliminated when applying CKM unitarity. However, a consequence of
Eq.~\eqref{eqn:CCR} is that the off-diagonal, basis dependent terms in
Eq.~\eqref{eqn:OCP} do not appear in the physical relations, as desired. 

The full set of the invariants, $[\mathcal{M}_{\rm CP}]_{w,v}$, is shown on the
left side of Table \ref{tab:USCPV} for $B$ decays to charged mesons. From the
general construction of CP sum rules in Eq.~\eqref{eqn:SRGR}, one may derive
U-spin CP sum rules in the basis of amplitudes $\{\mathcal{A}_{\mu \to
\alpha\beta}\}$ by solving $S^{\alpha\beta\mu}[\mathcal{M}_{\rm
CP}]_{w,v;\alpha\beta\mu} = 0$. That is, we need to find the null space of a
matrix whose entries are the invariant matrices $[\mathcal{M}_{\rm CP}]_{w,v;
\alpha\beta\mu}$ (see Ref.~\cite{Grossman:2012ry} for analogous amplitude and
rate sum rule constructions).  For the present analysis, the U-spin relations
may be read off Table~\ref{tab:USCPV}. For example, we have in the U-spin limit
$[\mathcal{M}_{\rm CP}]_{w,v;K^+\pi^- B_d} + [\mathcal{M}_{\rm
CP}]_{w,v;K^-\pi^+ B_d} = 0$, which immediately implies
\begin{equation}
	\label{eqn:DKP}
	\GAcp{B_d}{\pi ^- K^+ } + \GAcp{B_s}{\pi ^+ K^- } =0~.
\end{equation}
Explicitly,
\begin{align}
	[\mathcal{M}_{\rm CP}]_{w,v;K^+\pi^- B_d} = \begin{pmatrix} 0 & 0 \\ 0 & -J \end{pmatrix}, 
	\qquad
	[\mathcal{M}_{\rm CP}]_{w,v;K^-\pi^+ B_s} = \begin{pmatrix} 0 & 0 \\ 0 & +J \end{pmatrix},
\end{align}
and applying Eq.~\eqref{eqn:WEUD} yields
\begin{equation}
	\GAcp{ B_d}{\pi ^- K^+ } = 4J\, |X_1|^2|\mathcal{P\,T}| \sin\delta~,\qquad 
	\GAcp{ B_s}{\pi ^+ K^- } = -4J\, |X_1|^2|\mathcal{P\,T}| \sin\delta~.
\end{equation}

\begin{table}[t]
\begin{center}
\renewcommand{\arraystretch}{.9}
\begin{tabular*}{0.8\linewidth}{| @{\extracolsep{\fill}} >{$} c <{$} | >{$} c <{$} >{$}  c <{$} >{$}  c <{$}| >{$}  c <{$}  >{$} c <{$}  >{$}  c <{$} >{$}  c <{$} |}
\hline
\multirow{2}{*}{Decay mode}  &  \multicolumn{2}{c}{U-spin limit}  & &
  \multicolumn{3}{c}{U-spin breaking} & \\
&  \multicolumn{2}{c}{$ [\mathcal{M}_{\rm CP}]_{w,v} / J$}  & &
 \multicolumn{3}{c}{$ [\mathcal{M}_{\rm CP}]^{(0),(1)}_{w,v} / J$} & \\
\hline\hline
  \multirow{2}{*}{$~\GAcp{ B_s}{K^- \pi ^+ }~$}
& 0 & 0 & &0 & 0 & 0 & \\
& 0 & 1 & &0 & 0 & \epsilon  
  & \\
\hline
  \multirow{2}{*}{$~\GAcp{ B_d}{\pi ^- K^+ }~$}
& 0 & 0 & &0 & 0 & 0 & \\
& 0 & -1 && 0 & 0 & \epsilon 
  & \\
\hline\hline
  \multirow{2}{*}{$~\GAcp{B_d}{\pi ^- \pi ^+ }~$}
& \frac{1}{4} & \frac{1}{4} & &\frac{\epsilon }{4} & \frac{\epsilon }{2} & \frac{\epsilon
   }{4} & \\
& \frac{1}{4} & \frac{1}{4} & &\frac{\epsilon }{4} & \frac{\epsilon }{2} & \frac{\epsilon
   }{4} 
  & \\
    \hline
  \multirow{2}{*}{$~\GAcp{ B_s}{K^- K^+ }~$}
& -\frac{1}{4} & -\frac{1}{4} & & \frac{\epsilon }{4} & \frac{\epsilon }{2} &
   \frac{\epsilon }{4} & \\
& -\frac{1}{4} & -\frac{1}{4} & &\frac{\epsilon }{4} & \frac{\epsilon }{2} &
   \frac{\epsilon }{4} & \\
  \hline\hline
  \multirow{2}{*}{$~\GAcp{ B_s}{\pi ^- \pi ^+ }~$}
& -\frac{1}{4} & \frac{1}{4} & & \frac{\epsilon }{4} & -\frac{\epsilon }{2} &
   -\frac{\epsilon }{4} & \\
& \frac{1}{4} & -\frac{1}{4} & &-\frac{\epsilon }{4} & \frac{\epsilon }{2} &
   \frac{\epsilon }{4}
  & \\
  \hline
  \multirow{2}{*}{$~\GAcp{ B_d}{K^- K^+ }~$}
& \frac{1}{4} & -\frac{1}{4} & &\frac{\epsilon }{4} & -\frac{\epsilon }{2} &
   -\frac{\epsilon }{4} & \\
& -\frac{1}{4} & \frac{1}{4} & &-\frac{\epsilon }{4} & \frac{\epsilon }{2} &
   \frac{\epsilon }{4}
  & \\
  \hline
\end{tabular*}
\caption{Invariant matrices describing CPV in the U-spin limit and at first
  order in U-spin breaking in $B_{d,s}$ decays to $K^\pm$ and $\pi^\pm$. The
  $2\times 2$ ($2\times 3$) blocks should be read as matrices in indices $w$ and $v$, that are
  multiplied on the left by leading order reduced matrix elements and on the
  right by leading (first order in U-spin breaking) conjugate reduced matrix
  elements to produce a contribution to the corresponding $\gAcp$. For example,
  the top $2\times 2$ block is multiplied on the left by $(X_{0}, X_{1})$ and on
  the right by $(X_{0},  X_{1})^\dagger$, so that $X_w X_v^*[\mathcal{M}_{\rm
  CP}]_{w,v;K^-\pi^+B_s} = J|X_1|^2$, whereas the top $2\times 3$ block is
  multiplied by $(X_{0},  X_{1})$ on left and $(X^{(1)}_{0}, 
  X^{(1)}_{11},X^{(1)}_{12} )^\dagger$ on the right  (see Eq.~\eqref{eqn:UBI}
  for definitions of the subscripts), so that it contributes $X_w
  X_v^*[\mathcal{M}^{(0),(1)}_{\rm CP}]_{w,v;K^-\pi^+B_s} = \varepsilon J   X_{1}
  X^{(1)*}_{12}$.}
\label{tab:USCPV}
\end{center}
\end{table}

In the U-spin limit, the phase space factors for both modes are the same, so
that Eq.~\eqref{eqn:DKP} is equivalent to $\DAcp{B_d}{\pi^- K^+ } +
\DAcp{B_s}{\pi^+ K^- } =0$, from which $\Delta =0$ follows immediately. The
current experimental data imply
\begin{equation}
	\label{eqn:DGF}
	\frac{\GAcp{ B_d}{\pi ^- K^+ }}{\GAcp{ B_s}{\pi ^+ K^- }} + 1 = -0.05 \pm 0.22~.
\end{equation}
Finally, one may also see from Table~\ref{tab:USCPV} that in the U-spin
limit, we have two other relations,
\begin{align}
   \GAcp{B_d}{ \pi ^- \pi ^+} + \GAcp{B_s}{ K^- K^+} & = 0\,,\notag\\ 
   \GAcp{B_d}{ K^- K^+} + \GAcp{B_s}{ \pi ^- \pi ^+} & = 0\,.\label{eqn:DPX}
\end{align}
These relations generate other $\Delta$-type parameters, discussed further in
Sec.~\ref{sec:BDR}.

\subsection{First Order U-spin Breaking}

U-spin breaking arises from the mass splitting between the $d$ and $s$ quarks,
and may be encoded in the effective Hamiltonian by an expansion in a strange
quark mass spurion. This spurion transforms as a U-spin triplet, with vacuum
expectation value
\begin{equation}
	[m_s]^i_j = \varepsilon \begin{pmatrix} 1 & 0 \\ 0 & -1 \end{pmatrix},
\end{equation}
where $\varepsilon$ parametrizes U-spin breaking.

The CP asymmetry result in Eq.~\eqref{eqn:WEUD} follows immediately from
Eqs.~\eqref{eqn:HUR} and \eqref{eqn:AFP}, which hold for arbitrary
numbers of spurion insertions in the Hamiltonian. Hence Eq.~\eqref{eqn:WEUD}
holds to arbitrary order in U-spin breaking by this spurion: for each insertion
we just gain more group theoretic invariants, indexed by $w$. In particular, the
first order, $\mathcal{O}(\varepsilon)$, effects arise from a single insertion
of this spurion into the effective Hamiltonian. This is equivalent to three new
U-spin breaking partial invariants
\begin{align}
	[C_{11,j}^{(1)}]_{\alpha\beta\mu}  & = \partial_{P_\alpha P_\beta B_\mu} \Big\{ [M_1]^i_k [m_s]^k_i [B]_j\Big\}~,\notag\\
	[C_{12,j}^{(1)}]_{\alpha\beta\mu} &  = \partial_{P_\alpha P_\beta B_\mu} \Big\{[B]_k [M_1]^k_i [m_s]^i_j \Big\}~,\notag\\
	[C_{0,j}^{(1)}]_{\alpha\beta\mu}   & =  \partial_{P_\alpha P_\beta B_\mu}\Big\{ [M_0] [B]_i [m_s]^i_j\Big\}~, \label{eqn:UBI}
\end{align}
the corresponding contributions to the amplitudes are
\begin{equation}
	\mathcal{A}^{(1)}_{\mu \to \alpha\beta} = \sum_{w = 0,11,12} X_w^{(1)} [C_{w,j}^{(1)}]_{\alpha\beta\mu} H^j~.
\end{equation}
Furthermore, since $[m_s]^i_j $ is diagonal, Eq.~\eqref{eqn:CCR} continues to
hold, so the unphysical off-diagonal terms in $\mathcal{O}_{\rm CP}$ do not
appear in CP asymmetries. As a consequence of this, and applying 
Eq.~\eqref{eqn:WEUD}, the first order in U-spin breaking contributions to
$\gAcp$ are
\begin{align}
	\GAcp{B_\mu}{P_\alpha P_\beta} 
	& = 	\sum_{w,v }\!\! \bigg\{ \mbox{Im}\Big[\!X^{*}_wX^{(1)}_ve^{i \delta}\!\Big][\mathcal{M}^{(0),(1)}_{\rm CP}]_{w,v; \alpha\beta\mu} + \mbox{Im}\Big[\!X^{*(1)}_vX^{}_w e^{i \delta}\!\Big][\mathcal{M}^{(1),(0)}_{\rm CP}]_{v,w; \alpha\beta\mu}\bigg\}~,\notag\\
	& =  8\sin\delta\, |\mathcal{P}\, \mathcal{T}| \mathop{\sum_{w = 0,1}}_{v = 0,11,12}\mbox{Re}\big[X^*_w X^{(1)}_v \big] [\mathcal{M}^{(0),(1)}_{\rm CP}]_{w,v; \alpha\beta\mu}  ~,\label{eqn:OECP}
\end{align}	
where
\begin{equation}
	[\mathcal{M}_{\rm CP}]^{(0),(1)}_{w,v;\alpha\beta\mu} = [C_{w,i}]_{\alpha\beta\mu} \mathcal{O}^{ij}_{\rm CP} [C^{(1)}_{v,j}]_{\alpha\beta\mu} = [\mathcal{M}_{\rm CP}]^{(1),(0)}_{v,w;\alpha\beta\mu}~.
\end{equation}
The latter equality holds, because only the diagonal components of
$\mathcal{O}_{\rm CP}$ are physical. The matrices $\mathcal{M}_{\rm
CP}^{(0),(1)}$ are shown on the right side of Table~\ref{tab:USCPV}.
Finally,  one can see that there are no relations for charged kaons and pions
which hold when first order U-spin breaking is included. 

\section{Better defined relations and predictions}
\label{sec:BDR}

\subsection{Natural Parameters}

We see from Eq~\eqref{eqn:DKP} that the $\wt\Delta$ parameter, defined in
Eq.~\eqref{eqn:DTD}, vanishes in the U-spin limit. In group
theoretic notation and in terms of phase space factors, the $\Delta$ parameter, quoted by LHCb and defined in
Eq.~\eqref{eqn:DEV}, has the form
\begin{align}
	\Delta 
	& =  \frac{\dr{B_s}{\pi ^+ K^- } }{\dr{ B_d}{\pi ^- K^+ } } 
	\bigg[\frac{\DAcp{ B_d}{\pi ^- K^+ }}{\DAcp{ B_s}{\pi ^+ K^- }} + 1 \bigg] \notag\\
	&= \frac{\dr{B_s}{\pi ^+ K^- } }{\dr{ B_d}{\pi ^- K^+ } } 
	\bigg[\frac{\ps{B_d}{\pi^- K^+}\, \GAcp{ B_d}{\pi ^- K^+ }}
	{\ps{B_s}{\pi^+ K^-}\, \GAcp{ B_s}{\pi ^+ K^- }} + 1 \bigg]~.
	\label{eqn:DEF}
\end{align}
As mentioned above, in the U-spin limit the phase space factors are the same, so $\Delta =0$, too.
Furthermore, Table~\ref{tab:USCPV} and Eqs.~\eqref{eqn:DPX} imply that there are two other $\Delta$-type U-spin
breaking parameters involving charged kaon and pion final states that vanish
in the U-spin limit. If one were to enforce an analogy with Eqs.~\eqref{eqn:DEV} and
\eqref{eqn:DEF}, these parameters should be similarly written in the form
\begin{align}
	\Delta' & \equiv \frac{\dr{B_d}{\pi ^+ \pi^-}}{\dr{B_s}{K^+ K^- }}
	\bigg[\frac{\DAcp{B_s}{K^+ K^-}}{\DAcp{B_d}{\pi ^+ \pi^-}} + 1 \bigg] \,,
	\label{eqn:Delpdef} \\[2pt]
	\Xi & \equiv \frac{\dr{B_s}{\pi^+ \pi^- } }{\dr{ B_d}{K^+ K^-}} 
	\bigg[\frac{\DAcp{B_d}{K^+ K^- }}{\DAcp{B_s}{\pi^+ \pi^-}} + 1 \bigg] 
	\,. \label{eqn:Xidef}
\end{align}
Diagrammatically, the decays in $\Xi$ only receive contributions involving
$W$-exchange, penguin annihilation, or rescattering, which are power suppressed
in the heavy quark limit.  The $B_s\to\pi^+\pi^-$ decay was observed
recently~\cite{LHCbtalk}, and its rate is probably much larger than $B_d\to
K^+K^-$, which has not yet been seen with more than $2\sigma$ significance.

The key point here is that the values of $\Delta$, $\Delta'$, and $\Xi$ are not only
determined by the U-spin breaking in the square amplitude relations
(Eq.~\eqref{eqn:DKP} and its analogs \eqref{eqn:DPX}), but also by  the ratios of decay rates
and phase space factors.  Such normalizations lead to additional enhancements
or suppressions, so we expect that
\begin{equation}
	\Delta \sim 2\, \varepsilon\, \frac{\dr{B_s}{K^-\pi^+}} {\dr{B_d}{\pi^-K^+}}~,\qquad 
	\Delta' \sim 2\, \varepsilon\, \frac{\dr{B_d}{\pi^+\pi^-}}{\dr{B_s}{K^+K^-}}~,\qquad 
	\Xi \sim 2\, \varepsilon\, \frac{\dr{B_s}{\pi^+\pi^-}} {\dr{B_d}{K^+K^-}}~. 
	\label{eqn:DDX}
\end{equation}
The factors of two arise from the Taylor expansion in $\varepsilon$, the amplitude level breaking, of relations between squared amplitudes. The branching ratios collected in Table~\ref{values} then imply that we should
in turn expect
\begin{equation}\label{expectation}
	\Delta = 2\, {\cal O}(\varepsilon/4)~,\qquad 
	\Delta' = 2\, \mathcal{O}(\varepsilon/5)~. 
\end{equation}
With a canonical magnitude for U-spin breaking, $\varepsilon \sim 0.2$, we then
expect $\Delta \sim 0.10$, in good agreement with the data shown in
Eq.~\eqref{eqn:DEV}. The recent first LHCb measurement, $\Acp{B_s}{K^+K^-} =
-0.14 \pm 0.11$~\cite{Aaij:2013tna} and the world averaged
$\Acp{B_d}{\pi^+\pi^-} = 0.30 \pm 0.05$~ \cite{Amhis:2012bh,Aaij:2013tna},
provides $\Delta' = -0.26 \pm 0.38$, which agrees with the expectation
$\Delta' \sim 0.08$.  (Note that Ref.~\cite{Aaij:2013tna} quotes
$C_{KK}$, which is $-\Acp{B_s}{K^+K^-}$, under the extra assumption $|q/p|=1$.)

These extra normalization and phase space factors render $\Delta$, $\Delta'$, and $\Xi$ somewhat
arbitrary parameters to characterize the magnitude of U-spin breaking.  A set of
better-defined quantities are
\begin{align}
\wt{\Delta}\phantom{'} & \equiv \frac{\GAcp{B_d}{K^+\pi^-} + \GAcp{B_s}{K^-\pi^+}}
  {\GAcp{B_d}{K^+\pi^-} - \GAcp{B_s}{K^-\pi^+}}\,, \label{Dtdef}\\[5pt]
\wt{\Delta}' & \equiv \frac{\GAcp{B_s}{K^+K^-} + \GAcp{B_d}{\pi^+\pi^-}}
  {\GAcp{B_s}{K^+K^-} - \GAcp{B_d}{\pi^+\pi^-}}\,, \label{Dtpdef}\\[5pt]
\wt{\Xi}\phantom{'} & \equiv \frac{\GAcp{B_d}{K^+K^-}+\GAcp{B_s}{\pi^+\pi^-}}
  {\GAcp{B_d}{K^+K^-}-\GAcp{B_s}{\pi^+\pi^-}} \,. \label{Xtdef}
\end{align}
In contrast with Eq.~(\ref{expectation}), U-spin breaking with its canonical
magnitude predicts
\begin{equation}
\label{expectation2}
\Delta \sim \Delta' \sim \Xi \sim {\cal O}(\varepsilon) \,.
\end{equation}
For this reason, it is more natural to consider these parameters to study U-spin
breaking. Recent data (see Table \ref{values}) provides 
\begin{equation}
	\label{eqn:CTDDP}
	\wt \Delta = 0.026 \pm 0.106~,\qquad \mbox{and} \qquad \wt{\Delta} ' = 0.40 \pm 0.34~.
\end{equation}
Both values are in agreement with U-spin breaking expectations.

\begin{table}[tb]
\tabcolsep 16pt
\renewcommand{\arraystretch}{.9}
\begin{tabular}{|c|c|}
\hline
Parameter  &  Value \\
\hline\hline
$\Acp{B_s}{K^-\pi^+}$  &  $0.27 \pm 0.04$ \cite{Aaij:2013iua} \\
$\Acp{B_d}{K^+\pi^-}$  &  $-0.080 \pm 0.0076$ \cite{Amhis:2012bh} \\
$\Acp{B_s}{K^-K^+}$ & $-0.14 \pm 0.11$ \cite{Aaij:2013tna}\\
$\Acp{B_d}{\pi^-\pi^+}$ & $0.30 \pm 0.05$ \cite{Amhis:2012bh,Aaij:2013tna}\\
\hline
${\cal B}[B_d\to K^+\pi^-]$  &  $(19.55 \pm 0.54) \times 10^{-6}$ \cite{Amhis:2012bh} \\
${\cal B}[B_s\to K^-\pi^+]$  &  $(5.4 \pm 0.6) \times 10^{-6}$ \cite{Amhis:2012bh} \\
${\cal B}[B_s\to K^+K^-]$  &  $(24.5 \pm 1.8) \times 10^{-6}$ \cite{Amhis:2012bh} \\
${\cal B}[B_d\to \pi^+\pi^-]$  &  $(5.1 \pm 0.19) \times 10^{-6}$ \cite{Amhis:2012bh} \\
${\cal B}[B_s\to \pi^+\pi^-]$  &  $\big(0.95^{+0.25}_{-0.21}\big) \times 10^{-6}$ 
\cite{[] [{ \url{http://cds.cern.ch/record/1540696}.}] LHCbtalk} \\
${\cal B}[B_d\to K^+K^-]$  &  $(0.12\pm0.05) \times 10^{-6}$ \cite{Amhis:2012bh} \\
\hline
$\tau_{B_s}/\tau_{B_d}$  &  $0.998\pm0.009$ \cite{Amhis:2012bh} \\
\hline
$f_K/f_\pi$  &  $1.1936 \pm 0.0053$ \cite{[] [{, and updates at
\url{http://latticeaverages.org/}.}] Laiho:2009eu} \\
\hline
$\ps{B_d}{K^+\pi^-}$ & $1.066 \times 10^4$ MeV \cite{pdg:2012} \\
$\ps{B_s}{K^-\pi^+}$ & $1.083 \times 10^4$ MeV \cite{pdg:2012} \\
$\ps{B_d}{\pi^+\pi^-}$ & $1.058 \times 10^4$ MeV \cite{pdg:2012} \\
$\ps{B_s}{K^+K^-}$ & $1.091 \times 10^4$ MeV \cite{pdg:2012} \\
\hline
\end{tabular}
\caption{The numerical inputs used.}
\label{values}
\end{table}

\subsection{Heavy quark limit and factorization}

It has been shown that in the $m_b \gg \lqcd$ limit, the amplitudes of many $B$
decays to pairs of light mesons can be factorized into calculable short distance
factors: the $B\to X$ form factor, where meson $X$ inherits the (quantum numbers
of the) spectator quark in the $B$ meson, and the decay constant of the other
meson.  In all approaches to factorization~\cite{Bauer:2004tj, Beneke:2001ev,
Keum:2000wi}, the dominant amplitudes to the following decays can be written
at leading order in the form
\begin{align}
& \mathcal{A}(B_d \to K^+\pi^-) \propto F_{B_d \to \pi}\, f_K \,,
&& \mathcal{A}(B_s \to K^-\pi^+) \propto F_{B_s \to K}\, f_\pi \,,\notag\\
& \mathcal{A}(B_d \to \pi^+\pi^-) \propto F_{B_d \to \pi}\, f_\pi \,,
&& \mathcal{A}(B_s \to K^-K^+) \propto F_{B_s \to K}\, f_K\,.
\end{align}
However, there is limited agreement among different approaches to factorization
regarding the dominant source of strong phases, and the properties of
electroweak penguin, penguin annihilation, and $W$-exchange contributions
relative the leading terms.  

In the QCD factorization (BBNS)
approach~\cite{Beneke:2001ev, Beneke:2003zv} the dominant contributions to the
amplitudes with possibly large strong phases arise from power-suppressed
effects, which are modeled.  We find
\begin{align}
	\label{rel1}
	\Delta &\simeq \frac{\dr{B_s}{K^-\pi^+}} {\dr{B_d}{\pi^-K^+}}\,
        \bigg[ \bigg(\frac{F_{B_d \to \pi}}{F_{B_s \to K}}\,
        \frac{f_K}{f_\pi}\bigg)^2 - 1 \bigg]\,,\notag \\[5pt]
	\Delta'  &\simeq \frac{\dr{B_d}{\pi^+\pi^-}}{\dr{B_s}{K^+K^-}} \,
	\bigg[ \bigg(\frac{F_{B_s \to K}}{F_{B_d \to\pi}}\, 
	\frac{f_K}{f_\pi} \bigg)^2- 1 \bigg]\,,
\end{align}
or, in terms of the natural CP parameters, 
\begin{equation}\label{rel2}
\wt\Delta \simeq \frac{(F_{B_d \to \pi}\, f_K)^2
  - (F_{B_s \to K}\, f_\pi)^2}
  {(F_{B_d \to \pi}\, f_K)^2 + (F_{B_s \to K}\, f_\pi)^2} ~, \qquad
\wt\Delta' \simeq \frac{(F_{B_s \to K}\, f_K)^2 
  - (F_{B_d \to \pi}\, f_\pi)^2}
  {(F_{B_s \to K}\, f_K)^2 + (F_{B_d \to \pi}\, f_\pi)^2} ~.
\end{equation}
Here we used the simplified expressions adopted in Ref.~\cite{Beneke:2003zv},
and kept only the dominant source of direct CP violation proportional to
$\alpha_1\, \hat\alpha_4^c$, which is a good approximation numerically, since
$\beta_3^c$ is several times larger than $\beta_4^c$: see~\cite{Beneke:2003zv,
Beneke:2001ev} for definitions. (Similar results were also stated in Ref.~\cite{He:2013vta}.)
One may then eliminate the form factors from Eq.~(\ref{rel1}) to obtain
\begin{equation} \label{eqn:DDPR}
\Delta' \simeq \frac{\dr{B_d}{\pi^-\pi^+}}{\dr{B_s}{K^-K^+}}\, 
  \bigg[ \bigg(\frac{f_K}{f_\pi}\bigg)^4 
  \bigg( 1 + \Delta\, \frac{\dr{B_d}{\pi^-K^+}}{\dr{B_s}{K^-\pi^+}} \bigg)^{-1}
  - 1 \bigg] = 0.25 \pm 0.12 \,,
\end{equation}
where we used the numerical inputs collected in Table~\ref{values}. 
Alternatively, one can eliminate the form factors from Eq.~(\ref{rel2}) to
obtain,
\begin{equation}\label{eqn:DDPNR}
\wt{\Delta}' \simeq \bigg[ \bigg(\frac{f_K}{f_\pi}\bigg)^4\, 
  \frac{1-\wt\Delta}{1+\wt\Delta} -1 \bigg] \bigg/ 
  \bigg[\bigg(\frac{f_K}{f_\pi}\bigg)^4\, 
  \frac{1-\wt\Delta}{1+ \wt\Delta} + 1 \bigg] = 0.31 \pm 0.10\,,
\end{equation}
from the present value for $\wt\Delta$. These are in agreement with the recent LHCb measurements, that imply $\Delta' = -0.26 \pm
0.38 $ and $\wt{\Delta}' = 0.40 \pm 0.34$ respectively. The uncertainties are
expected to be reduced significantly in the future.

In the SCET (BPRS) approach~\cite{Bauer:2004tj, Bauer:2005kd} (see also \cite{Williamson:2006hb}) charm penguin
amplitudes are described as unsuppressed nonperturbative quantities, $A_{c\bar
c}^{M_1M_2}$, where $M_{1,2}$ are the final meson states, while other amplitudes with strong phases (relative to the leading
amplitudes) are ${\cal O}(\alpha_s, \lqcd/m_b)$.  If SU(3) breaking in the charm penguin amplitudes is small, then to a good approximation $A_{c\bar
c}^{K\pi} = A_{c\bar c}^{\pi\pi} = A_{c\bar c}^{KK}$~\cite{Bauer:2005kd}, so that one obtains Eq.~(\ref{rel2}) with all the squares removed.  Instead of Eq.~(\ref{eqn:DDPNR}), we obtain
\begin{equation}\label{eqn:DDPNRmod}
\wt{\Delta}' \simeq \bigg[ \bigg(\frac{f_K}{f_\pi}\bigg)^2\, 
  \frac{1-\wt\Delta}{1+\wt\Delta} -1 \bigg] \bigg/ 
  \bigg[\bigg(\frac{f_K}{f_\pi}\bigg)^2\, 
  \frac{1-\wt\Delta}{1+ \wt\Delta} + 1 \bigg] = 0.15 \pm 0.10\,.
\end{equation}
from the present value for $\wt\Delta$.
 
\begin{figure}[t!]
\centerline{\includegraphics[width=0.8\textwidth]{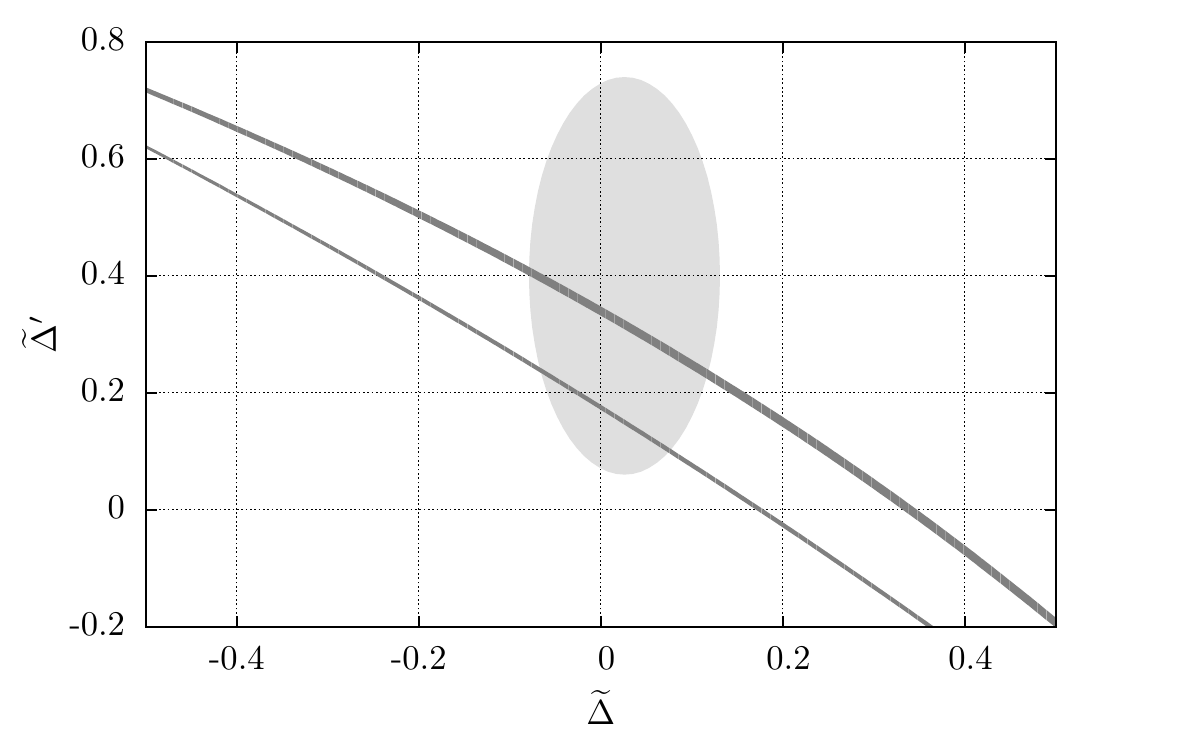}}
\caption{Factorization predictions for $\wt\Delta'$ as a function of
$\wt\Delta$. The upper [lower] gray bands show the prediction
Eq.~(\ref{eqn:DDPNR}) [Eq.~(\ref{eqn:DDPNRmod})]. Also shown is the present
$1\sigma$ confidence region for $\wt\Delta$ and $\wt{\Delta}'$ (gray ellipse)
assuming no experimental correlations. The widths of the bands indicate the
uncertainty from the lattice QCD calculation of $f_K/f_\pi$ (see
Table~\ref{values}).}
\label{sillyfig}
\end{figure}

The relations in Eqs.~\eqref{eqn:DDPNR} and \eqref{eqn:DDPNRmod} between
$\wt\Delta'$ and $\wt\Delta$ are displayed in Fig.~\ref{sillyfig}, compared with
present data for these parameters.  It shows that if factorization is a good
approximation then $\wt\Delta$ and $\wt\Delta'$ can only have comparable
magnitudes in a relatively small region.  In particular, if $\wt\Delta$ is close
to zero, as is its central value with the current data, then $\wt\Delta'$ should
deviate from zero substantially if subleading corrections to factorization are
small.  We see in Fig.~\ref{sillyfig} that this factorization picture conforms
with the current data, with the relations in Eqs.~\eqref{eqn:DDPNR} and \eqref{eqn:DDPNRmod} both intersecting
the present $1\sigma$ confidence region for $\wt\Delta$ and $\wt\Delta'$. In
contrast, observe that the U-spin limit prediction $(\wt\Delta, \wt\Delta') =
(0,0)$ does not agree as well with the current data; as already shown in 
Eqs.~\eqref{expectation2} and \eqref{eqn:CTDDP}, the prediction including first
order U-spin breaking effects is in concordance with the data. Future
comparisons of $\wt\Delta$ and $\wt\Delta'$ with these relations will probe the
factorization picture with greater precision.  Note that no serious lattice QCD
calculation of the $B_s \to K$ form factor exists yet, and these tests of
factorization should increase the motivations for such a calculation (besides
the hope of measuring $|V_{ub}|$ at LHCb from $\bar B_s\to K^+ e\bar\nu$).

Due to the lack of leading order contributions to the amplitudes in $\wt\Xi$ in
the heavy quark limit, and the complexity of the contributing power-suppressed
terms (see also Ref.~\cite{Arnesen:2006vb} and referenecs therein), this U-spin
relation may be expected to receive larger corrections, and $\wt\Xi$ is expected
to deviate from zero more significantly than $\wt\Delta$ and $\wt\Delta'$.  If
$\wt\Xi$ is measured in the future to be comparably close to zero as $\wt\Delta$
or $\wt\Delta'$, that would be a success of SU(3) flavor symmetry, and be
puzzling from the point of view of the heavy quark limit.

\section{SU(3) Relations}
\label{sec:SU3R}

Let us now proceed to consider full SU(3) and the CP relations that hold to
zeroth and first order in SU(3) breaking. We do not make any assumptions about
the size of the hadronic reduced matrix elements (see, e.g.,
Ref.~\cite{Chiang:2006ih} for such studies). However, we do make one assumption
that goes beyond flavor SU(3): we only consider effects that are first order in
the weak interaction.  In practice, this amounts to neglecting electroweak
penguin operators and $b\to d d \bar s$-type decays. This is well-justified as
corrections from higher order weak interactions corrections are expected to be
smaller than those from the SU(3) breaking induced by the quark mass spurion. 

\subsection{\texorpdfstring{\boldmath $B \to PP$}{BPP}}
\label{sec:BPP}

We consider first $B$ decays to two pseudoscalars. The initial states
furnish a flavor anti-triplet, and the final states an octet and singlet
\begin{equation}
	\label{eqn:BPVM}
	[B_3]_i = \begin{pmatrix} B^+ & B_d & B_s \end{pmatrix},\qquad 
	[P_1] = \eta_1, \qquad 
	[P_8]^i_j  = \begin{pmatrix} \dfrac{\pi^0}{\sqrt2} + \dfrac{\eta_8}{\sqrt6} & \pi^+ & K^+ \\ 
	\pi^- & - \dfrac{\pi^0}{\sqrt{2}} + \dfrac{\eta_8}{\sqrt6}  & K^0 \\ 
	K^- & \bar{K}^0 & -\dfrac{2\eta_8}{\sqrt6}\end{pmatrix}.
\end{equation}
The effective Hamiltonian is a four-quark current-current tensor operator,
\begin{equation}
	\label{eqn:EHTF}
	\mathcal{H}^{ij}_k =  (q^i \bar{q}_k)(q^j \bar{b})~,\qquad \mbox{or} \qquad \mathcal{H}^i = (q'\bar{q}')(q^i \bar{b})~,
\end{equation}
in which $q^i = (u,d,s)^T$ and $q' = c,b$.  The terms corresponding to charmless decays
transform as $\bm{3}\otimes\bar{\bm{3}}\otimes\bm{3} = \bm{3} \oplus \bm{3}'
\oplus \bar{\bm{6}} \oplus \bm{15}$. Enforcing charge conservation together
with CKM unitarity, the non-zero, independent components of each irrep are
\begin{align*}
 [\bm{3}]^2  & \simeq \frac{3}{2}\big[\mathcal{X}V^*_{cb}V_{cd} + \mathcal{Y}V^*_{ub}V_{ud}\big] 
 ~, \notag\\
 [\bm{3}]^3 & \simeq \frac{3}{2}\big[\mathcal{X}V^*_{cb}V_{cs} + \mathcal{Y}V^*_{ub}V_{us}\big]
 ~,\notag\\
 [\bm{3}']^2 & \simeq \frac{1}{2}\big[V^*_{ub}V_{ud} +  \mathcal{X}V^*_{cb}V_{cd} + \mathcal{Y}V^*_{ub}V_{ud}\big] 
 ~,\notag\\
 [\bm{3}']^3 & \simeq \frac{1}{2}\big[V^*_{ub}V_{us} +  \mathcal{X}V^*_{cb}V_{cs} + \mathcal{Y}V^*_{ub}V_{us}\big]
 ~,
\end{align*}
\begin{align}
& [\bar{\bm{6}}]_{12} \simeq \frac{1}{4}V^*_{ub}V_{us} 
~, && [\bar{\bm{6}}]_{13}\simeq -\frac{1}{4}V^*_{ub}V_{ud} 
~,\notag\\
& [\bm{15}]^{22}_2 \simeq -\frac{1}{4}V^*_{ub}V_{ud} 
~,  &&
[\bm{15}]^{23}_3 \simeq -\frac{1}{8}V^*_{ub}V_{ud} 
~,\notag\\  
&[\bm{15}]^{33}_3  \simeq -\frac{1}{4}V^*_{ub}V_{us} 
~, && [\bm{15}]^{32}_2  \simeq -\frac{1}{8}V^*_{ub}V_{us} 
~. \label{eqn:SU3EH}
\end{align}
Here $\mathcal{X}$ and $\mathcal{Y}$ are $\mathcal{O}(1)$ complex
numbers. As already mentioned we work to
first order in the weak interaction and thus in
Eqs.~\eqref{eqn:SU3EH} we have neglected electroweak penguin operators
as well as operators of the form $(q\bar{q}') (d\bar{b})$ with $q \ne
q'$. It is this assumption that is responsible for the fact that the
$\bar{\bm{6}}$ and $\bm{15}$ do not have terms proportional to $V^*_{cb}V_{cd}$.
Note that the extra penguin operator $(q'\bar{q}')(q^i \bar{b})$ in
\eqref{eqn:EHTF} furnishes a triplet proportional to the $\bm{3}$, and is
therefore subsumed by Eqs.~\eqref{eqn:SU3EH}.

Applying the Wigner-Eckart theorem, as in Sec.~\ref{sec:CPSR} and in particular
Eq.~\eqref{eqn:CPGR}, and assuming an arbitrary mixing angle between the $\eta$
and $\eta'$ mass eigenstates, we now present all possible SU(3) CP relations.
The first two are due to isospin
\begin{subequations}
\label{eqn:GLS}
\begin{align}
\GAcp{B_s}{ \pi^- \pi^+} &= 2\GAcp{B_s}{ 2\pi^0} \,, \label{eqn:IR1}\\ 
2\GAcp{B^+}{ \pi^0 K^+} -\GAcp{B^+}{ \pi^+ K^0}& 
  =  \GAcp{B_d}{ \pi ^- K^+} - 2\GAcp{B_d}{\pi^0 K^0} \,.\label{eqn:IR2} 
\end{align}
The next eight use only U-spin, of which the first three are the familiar
charged meson relations from Sec.~\ref{sec:UA},
\begin{align}
	\GAcp{B_d}{ \pi ^- K^+} & + \GAcp{B_s}{ \pi ^+ K^-} \label{eqn:GL1}= 0 \,,\\ 
	\GAcp{B_d}{ \pi ^- \pi ^+} & + \GAcp{B_s}{ K^- K^+} \label{eqn:GL2} = 0 \,,\\ 
  	 \GAcp{B_d}{ K^- K^+}& + \GAcp{B_s}{ \pi ^- \pi ^+}  \label{eqn:GL3}= 0 \,,\\ 
   	\GAcp{B_s}{ \pi^0 \bar{K}^0}&+\GAcp{B_d}{ \pi^0K^0} \label{eqn:GL4} = 0 \,,\\   
  	 \GAcp{B^+}{ K^+ \bar{K}^0} & + \GAcp{B^+}{\pi ^+ K^0} \label{eqn:GL5}= 0 \,,\\ 
  	 \GAcp{B_d}{ K^0 \bar{K}^0} &+ \GAcp{B_s}{ K^0  \bar{K}^0} = 0 \,,\label{eqn:GL6}\\
  	  \GAcp{B_s}{ \eta  \bar{K}^0}& +\GAcp{B_d}{ \eta K^0} = 0 \,, \label{eqn:SU31}  \\
  	 \GAcp{B_s}{ \eta' \bar{K}^0}& +\GAcp{B_d}{ \eta'  K^0} = 0\,, \label{eqn:SU32}
\end{align}
and the last two require full SU(3)
\begin{align}
   \GAcp{B^+}{ \pi ^+ \eta' }+\GAcp{B^+}{  \pi ^+ \eta }&+\GAcp{B^+}{ \eta'  K^+}\nn\\
   +\GAcp{B^+}{ \eta K^+} &+\GAcp{B^+}{ \pi^0 K^+}=0 \,,\label{eqn:SU33}\\[3pt]
   \GAcp{B_d}{ 2\eta' }+\GAcp{B_d}{ \eta'  \eta }& + \GAcp{B_d}{ \pi^0 \eta' }+\GAcp{B_d}{ 2\eta }\nn\\
   + \GAcp{B_d}{ \pi^0\eta }+\GAcp{B_d}{ 2\pi^0}& +\GAcp{B_s}{ 2\eta' }+ \GAcp{B_s}{ \eta' \eta}\nn\\
   +\GAcp{B_s}{ 2\eta } & +\GAcp{B_s}{ 2\pi^0}=0 \,.\label{eqn:SU34}
\end{align}
\end{subequations} 
The six relations \eqref{eqn:GL1}--\eqref{eqn:GL6} correspond to those found in
Refs.~\cite{He:1998rq, Gronau:2000zy}, while the first two relations,
\eqref{eqn:IR1}  and \eqref{eqn:IR2}, are isospin relation previously presented
in Refs.~\cite{Gronau:2005kz, Gronau:2008gu}. The relations
\eqref{eqn:SU31}--\eqref{eqn:SU34} are, to our knowledge, novel to this work. 
(For completeness, in Appendix~\ref{app:unmixed} we present relations in the
flavor basis. One may check these are consistent with the SU(3) decompositions
contained in Ref.~\cite{Grossman:2003qp}.) It should be noted that we have
chosen to present these sum rules in a particular basis, such that the U-spin
and isopsin sum rules are manifest. Of course, any linear combination of these
sum rules is also a sum rule.

Let us now incorporate $K^0$--$\bar{K}^0$ mixing. Since we work to first order
in the weak interaction, we neglect CPV effects in $K^0$--$\bar{K}^0$ mixing and
also neglect operators which produce $b\to d\bar sd$ - type decays.  Within
this approximation,  for each non-zero mode $B \to K^0 X$ ($B \to \bar{K}^0 X$)
the corresponding conjugate mode $B \to \bar{K}^0 X$ ($B \to K^0 X$) is zero.
Here $X$ denotes all pseudoscalar mesons with correct charges and $B$ denotes
$B_d$, $B_s$, or $B^+$ as appropriate.
It follows that 
\begin{equation}
	\label{eqn:KKM}
	\GAcp{B }{K_S X} = \begin{cases} \frac{1}{2}\, \GAcp{B}{K^0 X}~,\qquad \mbox{or} \\  \frac12\, \GAcp{B}{\bar K^0 X}~, \end{cases}
\end{equation}
for all pseudoscalar mesons $X$.  One further obtains the following relations 
\begin{subequations}
\label{eqn:KGLS}
\begin{align}
	\GAcp{B}{K_S X} & = \GAcp{B}{K_L X} \,, \label{eqn:KR1} \\
	\GAcp{B_{d,s}}{2K_S} & = \GAcp{B_{d,s}}{2K_L} \,, \label{eqn:KR3}
\end{align}
for all pseudoscalar mesons $X \not= K_{S,L}$ with correct charges. Note that
these relations arise from the properties of $K_{S,L}$, rather than from SU(3)
symmetry. 

In order to rotate to the $K$ meson mass basis,  we see from Eq.~\eqref{eqn:KKM}
that we need only replace $K^0$ and $\bar{K}^0$ in each of Eqs.~\eqref{eqn:GLS}
by $K_S$ (or $K_L$), with an extra factor of two in front of the corresponding
$\gAcp$.  Thus, in the mass basis we obtain the isospin relation
\begin{equation}
  	2\,\GAcp{B^+}{ \pi^0 K^+} -2\,\GAcp{B^+}{ \pi^+ K_{S}} 
	=\GAcp{B_d}{\pi ^- K^+} - 4\,\GAcp{B_d}{ \pi^0 K_{S}}\,, \label{eqn:KIR2}
\end{equation}
and the U-spin relations
\begin{align}
	\GAcp{B_d}{K_{S} X^0}  + \GAcp{B_s}{K_{S} X^0} & = 0\,, \label{eqn:KR2}\\
	\GAcp{B_d}{2K_{S}}   + \GAcp{B_s}{2K_{S}} & = 0\,, \label{eqn:KR4}\\
  	\GAcp{B^+}{ K^+K_{S}}   + \GAcp{B^+}{\pi ^+ K_{S}} & =0\,, \label{eqn:KGL4} 
\end{align}
\end{subequations} 
where $X^0 = \pi^0$, $\eta$, or $\eta'$.  The six relations in Eqs.~\eqref{eqn:GLS} not involving $K^0$ or $\bar{K}^0$
remain unchanged. 

All the above relations \eqref{eqn:GLS} or \eqref{eqn:KGLS}, once properly
normalized, are expected to receive corrections at $\mathcal{O}(\varepsilon)$
from SU(3) breaking. 
To compute CP relations that hold up to $\mathcal{O}(\varepsilon^2)$
corrections, one expands in the strange quark mass spurion, represented by
\begin{equation}
	\label{eqn:SQMS}
	m_s = \varepsilon\begin{pmatrix} ~1 & ~0 & ~0 \\ ~0 & ~1 & ~0 \\ ~0 & ~0 & -2\end{pmatrix}~.
\end{equation}
The isospin relations should hold to all orders in breaking by $m_s$ -- $m_s$
does not further break isospin --  but are clearly sensitive to isospin
breaking. In the flavor basis, we find that the isospin relations
\eqref{eqn:IR1} and \eqref{eqn:IR2}, together with
\begin{equation}
	\label{eqn:PPSO}
	 \GAcp{B_s }{ \pi^0 \bar{K}^0} -3\GAcp{B_s }{ \eta_8  \bar{K}^0} = 3 \GAcp{B_d }{ \eta_8  K^0} - \GAcp{B_d }{ \pi^0 K^0}
\end{equation} 
hold to $\mathcal{O}(\varepsilon^2)$. Furthermore, we find that
Eq.~\eqref{eqn:PPSO} holds to second order in isospin breaking. This is
result is novel to this work: while such a CP relation is untestable, it is
interesting to note such a relation exists in principle, given the large number
of cancellations required among the group theoretic invariants.

In the presence of $\eta$--$\eta'$ mixing, we find that only the isospin
relations \eqref{eqn:IR1} and \eqref{eqn:IR2} hold to
$\mathcal{O}(\varepsilon^2)$. If one includes $K^0$--$\bar{K}^0$ mixing, then
the relations \eqref{eqn:IR1} and  \eqref{eqn:KIR2} hold to
$\mathcal{O}(\varepsilon^2)$, along with the mixing relations \eqref{eqn:KR1}
and \eqref{eqn:KR3}, which do not arise from SU(3). Once isospin breaking is
introduced, there exists no CP relation that survives at first order.  In
summary, the $B \to PP$ isospin relations \eqref{eqn:IR1} and \eqref{eqn:KIR2}
are expected to hold to the $\mathcal{O}(1\%)$ level, while all other mass basis
CP relations should fail at $\mathcal{O}(\varepsilon)$.

\subsection{\texorpdfstring{\boldmath $B \to PV$}{BPV}}

We may also derive CP relations for charmless two-body $B$ decays to a
pseudoscalar and a vector meson. It should be noted that, experimentally, these
decays are measured via construction of Dalitz plots, and it is not always 
possible to identify the $PV$ final state. 

The vector mesons furnish an SU(3) singlet and octet, 
\begin{equation}
[V_1] = \phi_1 \,, \qquad
[V_8]^i_j   = \begin{pmatrix} \dfrac{\rho^0}{\sqrt2} + \dfrac{\omega_8}{\sqrt6} & \rho^+ & K^{*+} \\ 
  \rho^- & - \dfrac{\rho^0}{\sqrt2} + \dfrac{\omega_8}{\sqrt6}  & K^{*0} \\ 
  K^{*-} & \bar{K}^{*0} & - \dfrac{\omega_8}{\sqrt6} \end{pmatrix} ,
\end{equation}
the $B$ and pseudoscalars furnish the same representations as in
Eq.~\eqref{eqn:BPVM}. The effective Hamiltonian \eqref{eqn:SU3EH} and
strange quark spurion \eqref{eqn:SQMS} are unchanged.

Assuming ideal mixing between the $\omega$ and $\phi$ mass eigenstates, such
that $\phi$ is pure $s\bar{s}$, and arbitrary mixing between $\eta$ and $\eta'$,
one finds eighteen relations, corresponding to the following zero sums. This
first three are isospin relations
\begin{subequations}
\label{eqn:PVRLO}
\begin{align}
   2\GAcp{B_s}{ \pi^0 \rho^0}= \GAcp{B_s&}{ \rho ^- \pi ^+}+ \GAcp{B_s}{ \pi ^- \rho ^+}\,, \label{eqn:PVI1}\\
   \!2 \GAcp{B^+}{ \pi^0 K^{*+}}-\GAcp{B^+}{ \pi ^+ K^{*0}}& =
   \GAcp{B_d}{ \pi ^-K^{*+}}-2 \GAcp{B_d}{ \pi^0 K^{*0}}\,, \label{eqn:PVI3} \\
   2 \GAcp{B^+}{ K^+ \rho^0}-\GAcp{B^+}{ K^0 \rho ^+}& = \GAcp{B_d}{ K^+ \rho ^-} -2 \GAcp{B_d}{ K^0 \rho^0}\,. \label{eqn:PVI2}
\end{align}
The next ten are generated by U-spin
\begin{align}
       \GAcp{B_d}{ K^0 \bar{K}^{*0}}+\GAcp{B_s}{ K^{*0} \bar{K}^0} &= 0\,,\\ 
   \GAcp{B_d}{ K^{*0} \bar{K}^0}+\GAcp{B_s}{ K^0 \bar{K}^{*0}} &= 0\,,\\
   \GAcp{B_d}{ K^-K^{*+}}+\GAcp{B_s}{ \pi ^- \rho ^+} &= 0\,,\\
   \GAcp{B_d}{ \pi ^- \rho ^+}+\GAcp{B_s}{ K^- K^{*+}} &= 0\,,\\
   \GAcp{B_d}{ \pi ^- K^{*+}}+\GAcp{B_s}{ K^-\rho ^+} &= 0\,,\\
   \GAcp{B_d}{ K^+ \rho ^-}+\GAcp{B_s}{ \pi ^+ K^{*-}} &= 0\,,\\
   \GAcp{B_d}{ K^{*-} K^+}+\GAcp{B_s}{ \pi ^+ \rho ^-} &= 0\,,\\
   \GAcp{B_d}{ \pi ^+ \rho^-}+\GAcp{B_s}{ K^{*-} K^+} &= 0\,,\\
   \GAcp{B^+}{ K^{*+} \bar{K}^0}+\GAcp{B^+}{ K^0 \rho ^+} &= 0\,,\\
   \GAcp{B^+}{ K^+ \bar{K}^{*0}}+\GAcp{B^+}{ \pi ^+ K^{*0}} &= 0\,.
\end{align}
Finally, there are a further five SU(3) relations
\begin{align}
	\GAcp{B^+}{ \eta'  \rho ^+}+\GAcp{B^+}{ \eta  \rho ^+}&+\GAcp{B^+}{\pi^0 \rho ^+}\nn\\
	 +\GAcp{B^+}{ \eta'  K^{*+}}&+\GAcp{B^+}{ \eta  K^{*+}}+\GAcp{B^+}{ \pi^0 K^{*+}} = 0\,,\\[3pt]
	\GAcp{B^+}{ \pi ^+ \rho^0}+\GAcp{B^+}{ \pi ^+ \omega }&+\GAcp{B^+}{ \pi ^+ \phi }\nn\\
	 +\GAcp{B^+}{ K^+ \rho^0}&+\GAcp{B^+}{ K^+ \omega }+\GAcp{B^+}{ K^+ \phi } = 0\,,\\[3pt]
	\GAcp{B_s}{ \eta'  \bar{K}^{*0}}+\GAcp{B_s}{ \eta  \bar{K}^{*0}}&+\GAcp{B_s}{ \pi^0 \bar{K}^{*0}}\nn\\
	 +\GAcp{B_d}{ \eta'  K^{*0}}&+\GAcp{B_d}{ \eta  K^{*0}}+\GAcp{B_d}{ \pi^0 K^{*0}} = 0\,,\\[3pt]
	\GAcp{B_s}{ \rho^0 \bar{K}^0}+\GAcp{B_s}{ \omega  \bar{K}^0}&+\GAcp{B_s}{ \phi  \bar{K}^0}\nn\\
	 +\GAcp{B_d}{ K^0 \rho^0}&+\GAcp{B_d}{ K^0 \omega }+\GAcp{B_d}{ K^0 \phi }= 0\,,\\[3pt]
	\GAcp{B_d}{ \eta'  \rho^0}+\GAcp{B_d}{ \eta'  \omega }&+\GAcp{B_d}{ \eta'  \phi }+\GAcp{B_d}{ \eta  \rho^0}\nn\\
	 +\GAcp{B_d}{ \eta  \omega }+\GAcp{B_d}{ \eta  \phi}&+\GAcp{B_d}{ \pi^0 \rho^0}+\GAcp{B_d}{ \pi^0 \omega }\nn\\
	 +\GAcp{B_d}{ \pi^0\phi }+\GAcp{B_s}{ \eta'  \omega }&+\GAcp{B_s}{ \eta'  \phi }+\GAcp{B_s}{ \eta \omega }\nn\\
	 +\GAcp{B_s}{ \eta  \phi }&+\GAcp{B_s}{ \pi^0 \rho^0}= 0~.
\end{align}
\end{subequations}
Note that the $K^{*0}$ and $\bar{K}^{*0}$ can be tagged, so we need not consider
$K^*$--$K^*$ mixing. (As for the $B \to PP$ case, in Appendix~\ref{app:unmixed}
we present relations in the flavor basis. These are consistent with the SU(3)
decompositions contained in Ref.~\cite{Grossman:2003qp}.) Including
$K^0$--$\bar{K}^0$ mixing with the same approximations as in Sec.~\ref{sec:BPP},
leads to a further twelve relations
\begin{equation}
	\GAcp{B}{K_S X} = \GAcp{B}{K_L X}~,
\end{equation}
that do not arise from SU(3). Just as for the PP case, Eq.~\eqref{eqn:KKM} holds for all vector mesons $X$, so that the SU(3) CP relations for kaon mixing are obtained by
replacing all the $K^0$ and $\bar{K}^0$ mesons in Eqs.~\eqref{eqn:PVRLO} with
$K_S$, and including an extra factor of two in front of the corresponding
$\gAcp$. In particular, Eq.~\eqref{eqn:PVI2} becomes
\begin{equation}
	\label{eqn:KPVI2}
	2 \GAcp{B^+}{ K^+ \rho^0}-2\GAcp{B^+}{ K_S \rho ^+} = \GAcp{B_d}{ K^+ \rho ^-} -4 \GAcp{B_d}{ K_S \rho^0}~.
\end{equation}
We find that only the three isospin CP relations,
Eqs.~\eqref{eqn:PVI1}--\eqref{eqn:PVI2}, hold to second order in SU(3) breaking
by the strange quark mass spurion, with or without $\eta-\eta'$ mixing.
Including kaon mixing, the relation \eqref{eqn:PVI2} is replaced by
Eq.~\eqref{eqn:KPVI2}. Finally, with or without mixing, no CP relations hold at first order in isospin breaking. Similarly to $B \to PP$, we conclude that the three $B \to PV$ isospin relations  \eqref{eqn:PVI1}, \eqref{eqn:PVI3} and  \eqref{eqn:KPVI2} are expected to hold to the $\mathcal{O}(1\%)$ level, while all other mass basis CP relations should fail at $\mathcal{O}(\varepsilon)$.

\section{Conclusions}

New data on $B_s$ decays and CP asymmetries have made it possible to test
several U-spin and SU(3) relations. We have derived the complete set of leading order isospin, U-spin, and SU(3) CP relations, some of which are novel to this work. We further found that there are no relations for CP asymmetries
that hold at first order in SU(3) breaking, except for isospin relations. These
latter relations fail at first order in isospin breaking. While isospin relations are expected to hold at the percent level, this is not
the case with SU(3) relations, where the breaking effects are expected to be $\sim 20\%$. 

For the purposes of parametrizing SU(3) or U-spin breaking with these relations,
one must construct parameters that are properly normalized. Furthermore, the CP
relations themselves are formally constructed in terms of the phase
space-stripped decay rate splittings, $\gAcp$, which are well-defined in a group
theoretic sense, rather than in terms of the decay rate splittings, $\dAcp$.
Therefore, any such parameters that are designed to test the breaking of flavor
symmetries should be similarly constructed in terms of $\gAcp$, becuase they do
not admit extra breaking from phase space factors.

Factorization at leading order in the heavy quark limit predict relations
between the U-spin parameters $\wt\Delta$ and $\wt\Delta'$, given in
Eqs.~\eqref{eqn:DDPNR} and \eqref{eqn:DDPNRmod} and shown in Fig.~\ref{sillyfig}. We see that these factorization-based descriptions of U-spin breaking are in
good agreement with the data. We
hope that future data will test this picture with better precision.

From the flavor symmetry point of view, a third parameter $\wt\Xi$, defined in
Eq.~(\ref{Xtdef}), is on the same footing as $\wt\Delta$ and $\wt\Delta'$: we
expect corrections of $\mathcal{O}(\varepsilon)$. However, while the modes
relevant for $\wt\Delta'$ receive leading contributions in the heavy quark
limit, those in $\wt\Xi$ are power suppressed. Hence in the factorization
picture, $\wt\Xi$ may be expected to receive larger SU(3)-breaking corrections.
Thus, measurements of these parameters will help us understand which theoretical
tools are reliable.

In terms of future study, we have considered here only two-body decays. While
this is appropriate for decays into two pseudoscalars, $B \to PV$ decays are
measured through Dalitz analyses.  The Dalitz plots include the dominant
resonance regions, but also other features, such that a full study of the $B \to
3P$ decays would be well-motivated. 

\acknowledgments

We thank Gerhard Buchalla, Iain Stewart and Jure Zupan for helpful conversations.
While we were finalizing this work, Ref.~\cite{Gronau:2013mda}
appeared with U-spin results that partially overlap, and agree, with ours. The work of YG and DR  is supported by the U.S. National
Science Foundation through grant PHY-0757868 
and by the United States-Israel Binational Science
Foundation (BSF) under grant No.~2010221. The work of DR is also supported in part by the U.S. National
Science Foundation under Grant No.~PHY-1002399.
ZL~was supported in part by the Office of High Energy Physics of the U.S.\
Department  of Energy under contract DE-AC02-05CH11231.
ZL thanks the Erwin Schr\"odinger Institute program ``Jets and Quantum Fields
for LHC and Future Colliders" for hospitality while portions of this work were
completed.

\appendix

\section{CP Sum Rules in Flavor Basis}
 \label{app:unmixed}
In this appendix we present CP sum rules in the flavor basis, that is, without $K$--$\bar{K}$, or singlet-octet mixing. Clearly,
these cannot be tested experimentally, but we include them here for the sake of completeness.

There are 19 linearly independent $B \to PP$ sum rules in this basis, namely,
\begin{subequations}
\label{eqn:CPFPP}
\begin{align}
   2\GAcp{B_s}{ 2\pi^0}& =\GAcp{B_s}{ \pi ^- \pi ^+}\\   
   2 \GAcp{B^+}{ \pi ^0K^+}-\GAcp{B^+}{ \pi ^+ K^0}& =\GAcp{B_d}{ \pi ^- K^+}-2 \GAcp{B_d}{ \pi ^0 K^0}\\   
   \GAcp{B_d}{ \pi ^- \pi ^+}& +\GAcp{B_s}{ K^- K^+}=0\\   
   \GAcp{B_d}{ K^-K^+}& +\GAcp{B_s}{ \pi ^- \pi ^+}=0\\   
   \GAcp{B_d}{ K^0 \bar{K}^0}&+\GAcp{B_s}{ K^0\bar{K}^0}=0\\   
   \GAcp{B_d}{ \pi ^- K^+}& +\GAcp{B_s}{ \pi ^+ K^-}=0\\   
   \GAcp{B^+}{ K^+ \bar{K}^0}&+\GAcp{B^+}{ \pi ^+ K^0}=0\\   
   \GAcp{B_s}{ \pi ^0 \bar{K}^0}&+\GAcp{B_d}{\pi^0 K^0}=0\\   
   \GAcp{B_d}{ 2\eta_1}& +\GAcp{B_s}{ 2\eta_1}=0\\   
  \GAcp{B^+}{ \pi ^+\eta _1}& +\GAcp{B^+}{ \eta _1 K^+}=0\\   
   \GAcp{B_s}{ \pi ^0 \bar{K}^0}& =3 \GAcp{B_s}{\eta _8 \bar{K}^0}\\   
   \GAcp{B_s}{ \eta _1 \bar{K}^0}&+\GAcp{B_d}{ \eta _1 K^0}=0\\   
    \GAcp{B_s}{ \eta _8 \bar{K}^0}&+\GAcp{B_d}{ \eta _8 K^0}=0\\   
   \GAcp{B_d}{ \pi^0 \eta_1}+\GAcp{B_d}{ & \eta _1 \eta _8}+\GAcp{B_s}{ \eta _1 \eta _8}=0\\   
   \GAcp{B^+}{ \pi ^+ \eta _8}+\GAcp{B^+}{ & K^+\eta_8} + \GAcp{B^+}{ \pi^0 K^+}=0\\   
   \GAcp{B_s}{ \eta _1 \bar{K}^0}+\GAcp{B_d}{ \pi^0 \eta _1}& =\GAcp{B_d}{ \eta _1 \eta _8}-2 \GAcp{B_s}{ \eta _1 \eta _8}\\   
   \GAcp{B_s}{ \pi ^+ K^-}-\GAcp{B^+}{ K^+ \bar{K}^0}& =6 \GAcp{B_s}{ \eta _8 \bar{K}^0}+6 \GAcp{B^+}{ \eta _8 K^+}\\
   \GAcp{B_d}{\pi ^0 \eta _8}+\GAcp{B_d}{ 2\eta_8}+\GAcp{B_d}{& 2\pi^0}+\GAcp{B_s}{ 2\eta_8}+\GAcp{B_s}{ 2\pi^0}=0\\
   -\GAcp{B_d}{ K^0 \bar{K}^0}-2 \GAcp{B_s}{ \eta _8\bar{K}^0}&+\GAcp{B_s}{ K^0 \bar{K}^0}+2 \GAcp{B_d}{ \pi ^0 \eta _8}\notag\\
   +4 \GAcp{B_d}{ 2\eta_8}+2 \GAcp{B_d}{ \eta _8 K^0}&-2\GAcp{B_s}{ 2\eta_8}+2\GAcp{B_s}{ 2\pi^0}=0~.
\end{align}
\end{subequations}
Similarly, there are 32 linearly independent $B \to PV$ sum rules in the flavor basis, which are,
\begin{subequations}
\label{eqn:CPFPV}
\begin{align}
2 \GAcp{B_s}{ \pi^0  \rho^0  }=\GAcp{B_s}{ & \rho ^- \pi ^+}+\GAcp{B_s}{ \pi ^- \rho^+}\\   
   2 \GAcp{B_d}{ \pi^0  K^{*0} }-\GAcp{B_d}{ \pi ^- K^{*+}} & =\GAcp{B^+}{ K^{*0} \pi ^+}-2 \GAcp{B^+}{ \pi^0  K^{*+} }\\   
   2 \GAcp{B_d}{ K^0\rho^0}-\GAcp{B_d}{ \rho ^- K^+} & = \GAcp{B^+}{ K^0 \rho^+}-2 \GAcp{B^+}{ \rho^0  K^+ }\\   
   \GAcp{B^+}{ \bar{K}^{*0} K^+ } & +\GAcp{B^+}{ K^{*0}\pi ^+} = 0\\   
   \GAcp{B^+}{ K^0 \rho ^+} & +\GAcp{B^+}{ \bar{K}^0 K^{*+}} = 0\\   
   \GAcp{B^+}{ \eta_1  \rho ^+}& +\GAcp{B^+}{ \eta_1  K^{*+}} = 0\\   
   \GAcp{B_d}{ \rho ^- K^+ } & +\GAcp{B_s}{\bar{K}^{*-}\pi ^+} =0\\   
   \GAcp{B_d}{ \bar{K}^{*-}K^+ }& +\GAcp{B_s}{ \rho ^- \pi ^+} = 0\\   
   \GAcp{B_d}{ \rho ^- \pi ^+} & +\GAcp{B_s}{\bar{K}^{*-}K^+} = 0 \\   
   \GAcp{B_d}{ K^-K^{*+} } & +\GAcp{B_s}{ \pi ^- \rho ^+}  = 0\\   
   \GAcp{B_d}{ \pi ^- \rho ^+} & +\GAcp{B_s}{ K^- K^{*+} } = 0\\   
   \GAcp{B_d}{ \pi ^- K^{*+}} & +\GAcp{B_s}{ K^-\rho ^+} =0\\   
   \GAcp{B_d}{ K^0 \bar{K}^{*0}} & +\GAcp{B_s}{ \bar{K}^0 K^{*0}} = 0\\   
   \GAcp{B_d}{ \bar{K}^0 K^{*0} } & +\GAcp{B_s}{ K^0\bar{K}^{*0}} =0 \\   
   \GAcp{B_d}{ \omega_8 K^0 }+\GAcp{B_d}{ K^0 \rho^0  } & +\GAcp{B_s}{ \omega_8  \bar{K}^0}+\GAcp{B_s}{ \rho^0 \bar{K}^0} =0 \\   
   \GAcp{B^+}{ \omega_8  K^+}+\GAcp{B^+}{ \rho^0  K^+} & +\GAcp{B^+}{ \omega_8  \pi ^+}+\GAcp{B^+}{ \rho^0  \pi ^+} =0\\   
   +\GAcp{B^+}{ \omega_8  \pi ^+}+\GAcp{B^+}{ K^{*0} \pi^+} & = 2 \GAcp{B^+}{ \omega_8  K^+ } + \GAcp{B^+}{ \rho^0  \pi ^+}\\   
   \GAcp{B_d}{ \eta_8  K^{*0} }+\GAcp{B_d}{ \pi^0  K^{*0}} & +\GAcp{B_s}{ \eta_8 \bar{K}^{*0}}+\GAcp{B_s}{ \pi^0  \bar{K}^{*0}} =0 \\   
   \GAcp{B^+}{ \eta_8  \rho ^+}+\GAcp{B^+}{ \pi^0  \rho ^+} & +\GAcp{B^+}{ \eta_8  K^{*+}}+\GAcp{B^+}{ \pi^0  K^{*+}} = 0\\   
   \GAcp{B^+}{ \eta_8  \rho ^+}+\GAcp{B^+}{ K^0\rho ^+} &= 2 \GAcp{B^+}{ \eta_8  K^{*+}}+  \GAcp{B^+}{ \pi^0 \rho ^+}\\   
   \GAcp{B_d}{ \phi_1 K^0} & +\GAcp{B_s}{ \phi_1  \bar{K}^0} =0\\   
   \GAcp{B^+}{ \phi_1  K^+} & +\GAcp{B^+}{ \phi_1  \pi^+} =0\\   
   \GAcp{B_d}{ \eta_1  \phi_1 }& +\GAcp{B_s}{ \eta_1  \phi_1 } =0\\   
   \GAcp{B_d}{ \eta_1 K^{*0}}& +\GAcp{B_s}{ \eta_1  \bar{K}^{*0}}=0 \\   
   \GAcp{B_d}{ \eta_8  \phi_1 }+\GAcp{B_d}{ & \phi_1  \pi^0  }+\GAcp{B_s}{ \eta_8  \phi_1 } =0 \\   
   3 \GAcp{B_d}{ \eta_8  \phi_1 }+\GAcp{B_d}{ & \phi_1  K^0}+\GAcp{B_d}{ \phi_1  \pi^0  } =0\\   
   \GAcp{B_d}{ \eta_1  \omega_8 }+\GAcp{B_d}{ & \eta_1  \rho^0  }+\GAcp{B_s}{ \eta_1  \omega_8 } =0 \\   
   3 \GAcp{B_d}{ \eta_1  \omega_8 }+\GAcp{B_d}{ & \eta_1  \rho^0  }+\GAcp{B_d}{ \eta_1  K^{*0}} =0 
 \end{align}
 \begin{gather} 
	\begin{split}
   		\GAcp{B_d}{ \eta_8  \omega_8 }+\GAcp{B_d}{ \omega_8  \pi^0  }& +\GAcp{B_d}{ \eta_8  \rho^0  } \\
   			+\GAcp{B_d}{ \pi^0  \rho^0  }& +\GAcp{B_s}{ \eta_8  \omega_8 }+\GAcp{B_s}{ \pi^0  \rho^0 } = 0 
	\end{split} \\   
	\begin{split}
  		 \GAcp{B_d}{ \omega_8  K^0}+\GAcp{B_d}{ \omega_8  \pi^0   }& -\GAcp{B_d}{ \eta_8  \rho^0  } \\
		 	+\GAcp{B_d}{ \pi^0  K^{*0}} & +\GAcp{B_s}{ \rho^0  \bar{K}^0}+\GAcp{B_s}{ \eta_8  \bar{K}^{*0}} =0   
	\end{split}\\
	\begin{split}
 		 +\GAcp{B_d}{ \eta_8  \rho^0   }+\GAcp{B_d}{ K^0\rho^0  } & -\GAcp{B_d}{ \omega_8  \pi^0  }\\
		 	+\GAcp{B_d}{ \eta_8  K^{*0}}& +\GAcp{B_s}{ \omega_8  \bar{K}^0}+\GAcp{B_s}{ \pi^0  \bar{K}^{*0}} =0 
	\end{split}\\   
	\begin{split}
   		2 \GAcp{B_d}{ \omega_8  \pi^0  }-2 \GAcp{B_d}{ \eta_8  \rho^0  }& +\GAcp{B_d}{ \eta_8  K^{*0}}\\
			+\GAcp{B_d}{ \pi^0  K^{*0}} & +\GAcp{B_s}{ \omega_8 \bar{K}^0}+\GAcp{B_s}{ \rho^0  \bar{K}^0} =0 ~.
	\end{split}
\end{gather}
\end{subequations}

\section{Charm Decays}
Similarly to $B$ decays, we can compute CP relations for charmless $D \to PP$ and $D \to PV$. In this case, the initial states furnish an SU(3) triplet,  $[D]^i =   (~ D^0 ,~ D^+ ,~  D^+_s ~)^T$, and the 4-quark Hamiltonian has terms
\begin{equation}
	\label{eqn:CH}
	\mathcal{H}^{k}_{ij} = (\bar{q}_{i}q^k)(\bar{q}_j c)~, \qquad \mbox{or} \qquad \mathcal{H}_i = (\bar{c}c)(\bar{q}_i c)~.
\end{equation}
The charmless terms transform as
$\bar{\bm{3}}\otimes\bm{3}\otimes\bar{\bm{3}} = \bm{\bar{3}_{\rm p}} \oplus
 \bm{\bar{3}_{\rm t}} \oplus \bm{6} \oplus \bar{\bm{15}}$. Enforcing QED charge
conservation together with CKM unitarity, the non-zero, independent components of each irrep are
\begin{gather}
	[\bm{\bar{3}_p}]_1 \simeq -3\mathcal{X} V_{cb}^*V_{ub}~, \qquad  [\bm{\bar{3}_t}]_1 \simeq  \mathcal{X}V_{cb}^*V_{ub}~, \notag\\
	[\bm{6}]^{22}  \simeq \frac{1}{2}V_{cs}^*V_{ud}~, \qquad [\bm{6}]^{23}  \simeq -\frac{1}{4}\big(V_{cd}^*V_{ud} - V_{cs}^*V_{us}\big)~, \qquad [\bm{6}]^{33}  \simeq -\frac{1}{2}V_{cd}^*V_{us}~,\notag\\
	[\bm{\bar{15}}]^3_{12}  \simeq \frac{1}{2}V_{cd}^*V_{us}~, \qquad [\bm{\bar{15}}]^2_{13}  \simeq  \frac{1}{2}V_{cs}^*V_{ud}~, \notag\\
	[\bm{\bar{15}}]^2_{12}  \simeq \frac{3}{8}V_{cd}^*V_{ud} - \frac{1}{8}V_{cs}^*V_{us}~, \qquad [\bm{\bar{15}}]^3_{13}  \simeq \frac{3}{8}V_{cs}^*V_{us} - \frac{1}{8}V_{cd}^*V_{ud}~. \label{eqn:CKMH}
\end{gather}
Here $\mathcal{X}$ is an $\mathcal{O}(1)$ complex number. Penguin contributions carrying strong phases arise purely in the $\bar{\bm{3}}$ irreps, and note that the $[\bm{\bar{3}_t}]$ and $\bar{\bm{3}}$ irrep produced by the charm term in Eq.~\eqref{eqn:CH} are both subsumed by $[\bm{\bar{3}_p}]$. 

One finds, including $\eta$-$\eta'$ and  $K^0$-$\bar{K}^0$ mixing, the following leading order $D \to PP$ CP relations. There are two U-spin relations,
\begin{subequations}
\begin{align}
	 \GAcp{D^0}{ \pi ^- \pi ^+} & +\GAcp{D^0}{ K^-K^+}=0 \label{eqn:CKKPP}\\
	  \GAcp{D^+}{  K^+ K_S} & + \GAcp{D^+_s}{ \pi ^+ K_S}=0
\end{align}
two pure SU(3) relations,
\begin{gather} 
	 \begin{split}
  	 \GAcp{D^+}{ \pi ^+ \eta '}+\GAcp{D^+}{ \pi ^+ \eta }& +\GAcp{D^+_s}{ K^+ \eta '}\\
	 +\GAcp{D^+_s}{ \eta  K^+}& +\GAcp{D^+_s}{ \pi _0 K^+}=0
	 \end{split}\\[3pt]  
	\begin{split}
	\GAcp{D^0}{ 2\eta }+2 \GAcp{D^0}{ \eta  \eta '}&+2 \GAcp{D^0}{ \pi _0 \eta '}\\
		+\GAcp{D^0}{ 2\eta ' }&+2 \GAcp{D^0}{ \pi _0 \eta }+\GAcp{D^0}{2\pi ^0}=0~,
	\end{split}
\end{gather}
and the two mixing relations, that do not arise from SU(3)
\begin{align}
  \GAcp{D^+_s}{ \pi ^+ K_L}&= \GAcp{D^+_s}{ \pi ^+ K_S}\\  
   \GAcp{D^+}{ K^+ K_S}&=\GAcp{D^+}{ K^+ K_L}~.
\end{align}
\end{subequations}
(At first order in SU(3) breaking, there is also the mixing relation $
\GAcp{D^0}{ 2 K_S}=\GAcp{D^0}{ 2 K_L}$, each mode of which has zero direct CPV at
leading order.) The relation \eqref{eqn:CKKPP} is a well-known U-spin
relation~\cite{Grossman:2006jg}. No SU(3) relations hold at first order in breaking by the strange quark mass spurion or isospin breaking, with or without mixing.

Similarly, for $D \to PV$, we have, in the presence of neutral meson mixing, five U-spin relations
\begin{subequations}
\begin{align}
   \GAcp{D^0}{\pi ^- \rho ^+}&+\GAcp{D^0}{K^-K^{*+}} =0 \\
   \GAcp{D^0}{\pi^+ \rho ^-}&+\GAcp{D^0}{K^{*-} K^+}=0 \\
   \GAcp{D^+ }{ K^{*+} K_S}&+\GAcp{D^+_s}{\rho ^+ K_S}=0\\
   \GAcp{D^+}{K^+ \bar{K}^{*0}}&+\GAcp{D^+_s}{\pi ^+K^{*0}}=0 \\
   \GAcp{D^0}{\bar{K}^{*0} K_S}&+\GAcp{D^0}{K^{*0} K_S}=0~,
\end{align}
three SU(3) relations
\begin{gather} 
	\begin{split}
		 \GAcp{D^+}{\eta'  \rho ^+}-\GAcp{D^+}{\eta  \rho ^+}&-\GAcp{D^+}{\pi^0 \rho ^+}\\
		 	+\GAcp{D^+_s}{\eta' K^{*+}} &-\GAcp{D^+_s}{\eta K^{*+}}-\GAcp{D^+_s}{\pi^0K^{*+}} =0
	\end{split}\\[3pt]
	\begin{split}
		\GAcp{D^+}{\pi ^+ \rho^0}+\GAcp{D^+}{\pi ^+ \omega }&+\GAcp{D^+}{\pi ^+ \phi }\\
			+\GAcp{D^+_s}{K^+ \rho^0}&+\GAcp{D^+_s}{K^+ \omega }+\GAcp{D^+_s}{K^+ \phi } =0
	\end{split}\\[3pt]
	\begin{split}
		 \GAcp{D^0}{\eta'  \rho^0}+\GAcp{D^0}{\eta'  \omega }&+\GAcp{D^0}{\eta'  \phi}\\
		 -\GAcp{D^0}{\eta  \rho^0} &-\GAcp{D^0}{\eta  \omega }-\GAcp{D^0}{\eta  \phi}\\
		 -\GAcp{D^0}{\pi^0 \rho^0}&-\GAcp{D^0}{\pi^0 \omega }-\GAcp{D^0}{\pi^0 \phi } =0~,
	\end{split}
\end{gather} 
and four mixing relations
\begin{align}
   \GAcp{D^+_s}{ \rho ^+ K_S}&=\GAcp{D^+_s}{ \rho ^+ K_L}\\   
   \GAcp{D^+}{ K^{*+} K_S}&=\GAcp{D^+}{ K^{*+} K_L}\\   
   \GAcp{D^0}{ \bar{K}^{*0} K_S}&=\GAcp{D^0}{ \bar{K}^{*0} K_L}\\   
   \GAcp{D^0}{ K^{*0} K_S}&=\GAcp{D^0}{ K^{*0} K_L}~.
\end{align}
Once again, no SU(3) relations hold at first order in breaking by the strange quark mass spurion or isospin breaking, with or without mixing.
\end{subequations}


%

\end{document}